\documentclass[prd,preprint,showpacs,preprintnumbers,amsmath,amssymb,nofootinbib,superscriptaddress]{revtex4-1}
\usepackage{graphicx}
\usepackage{dcolumn}
\usepackage{bm}
\usepackage{subfigure}

\begin{document}

\preprint{UK/10-01}
\title{Overlap Valence on $2+1$ Flavor Domain Wall Fermion Configurations with Deflation and Low-mode Substitution} 
\collaboration{$\chi$QCD Collaboration}
\author{A. Li}
\affiliation{Dept.\ of Physics, Duke University, Durham, NC 27708}
\affiliation{Dept.\ of Physics and Astronomy, University of Kentucky, Lexington, KY 40506}
\author{A. Alexandru}
\affiliation{Dept.\ of Physics, George Washington University, Washington, DC 20052}
\author{Y. Chen}
\affiliation{Institute of High Energy Physics, Chinese Academy of Science, Beijing 100049, China}
\author{T. Doi}
\affiliation{Graduate School of Pure and Applied Science, University of Tsukuba, Tsukuba, Ibaraki 305-8571, Japan}
\author{S.J. Dong}
\affiliation{Dept.\ of Physics and Astronomy, University of Kentucky, Lexington, KY 40506}
\author{T. Draper}
\affiliation{Dept.\ of Physics and Astronomy, University of Kentucky, Lexington, KY 40506}
\author{M. Gong}
\affiliation{Dept.\ of Physics and Astronomy, University of Kentucky, Lexington, KY 40506}
\author{A. Hasenfratz}
\affiliation{Dept.\ of Physics, University of Colorado, Boulder, CO 80309}
\author{I. Horv\'{a}th}
\affiliation{Dept.\ of Physics and Astronomy, University of Kentucky, Lexington, KY 40506}
\author{F.X. Lee}
\affiliation{Dept.\ of Physics, George Washington University, Washington, DC 20052}
\author{K.F. Liu}
\affiliation{Dept.\ of Physics and Astronomy, University of Kentucky, Lexington, KY 40506}
\author{N. Mathur}
\affiliation{Department of Theoretical Physics, Tata Institute of Fundamental Research, Mumbai 40005, India}
\author{T. Streuer}
\affiliation{Institute for Theoretical Physics, University of Regensburg, 93040 Regensburg, Germany}
\author{J.B. Zhang}
\affiliation{ZIMP and Dept.\ of Physics, Zhejiang University, Hangzhou, Zhejiang 310027, China}

\begin{abstract}
The overlap fermion propagator is calculated on $2+1$ flavor domain wall fermion gauge configurations
on $16^3 \times 32$, $24^3 \times 64$ and $32^3 \times 64$ lattices. With HYP smearing and low eigenmode
deflation, it is shown that the inversion of the overlap operator can be expedited by $\sim 20$ times for the
$16^3 \times 32$ lattice and $\sim 80$ times for the $32^3 \times 64$ lattice. The overhead cost for calculating
eigenmodes ranges from 4.5 to 7.9 propagators for the above lattices.
Through the study of hyperfine splitting, we found that the $O(m^2a^2)$ error is small and
these dynamical fermion lattices can adequately accommodate quark mass up to the charm quark. A preliminary 
calculation of the low energy constant $\Delta_{mix}$ which characterizes the discretization error of the pion 
made up of a pair of sea and valence quarks in this mixed action approach is carried out via the scalar correlator 
with periodic and anti-periodic boundary conditions. It is found to be small which shifts a 300 MeV pion mass by 
$\sim 10$ to 19 MeV on these sets of lattices. We have studied the signal-to-noise issue of the noise source for 
the meson and baryon. We introduce a new algorithm with $Z_3$ grid source and low eigenmode substitution to study the 
the many-to-all meson and baryon correlators. It is found to be efficient in reducing errors for the correlators 
of both mesons and baryons. With 64-point $Z_3$ grid source and low-mode substitution, it can reduce the statistical 
errors of the light quark ($m_{\pi} \sim 200 - 300$ MeV) meson and nucleon correlators by a factor of $\sim 3-4$ as 
compared to the point source. The $Z_3$ grid source itself can reduce the errors of the charmonium correlators by a 
factor of $\sim 3$.
\end{abstract}

\maketitle
\section{Introduction}  \label{intro}

   A large scale endeavor has been undertaken by the RBC and UKQCD collaborations in the last few years to
simulate 2+1 flavor full QCD with dynamical domain wall fermions (DWF) and Iwasaki gauge action on several lattices
with pion mass as low as $\sim 300$ MeV and volume large enough for mesons 
($m_{\pi}L > 4$)~\cite{RBC-UKQCD07,RBC-UKQCD08,maw09}. Three sets of lattices
$16^3 \times 32 \times L_S$ and $24^3 \times 64 \times L_S$ at $a^{-1} = 1.73(3)$ GeV and $32^3 \times 64 \times L_S$
at $a^{-1} = 2.32(3)$ GeV with the fifth dimension $L_S = 16$ are available, each with 3 to 4 sea quark masses with
the lowest pion mass at $\sim 300$ MeV. With these
lattices, one can proceed to perform chiral extrapolation and continuum extrapolation assuming $a^2$ dependence
of the physical quantities. Since the domain wall fermion with $L_S = 16$ is a good approximation
for the chiral fermion satisfying Ginsparg-Wilson relation, it is shown that they have good chiral
properties and that most of the chiral symmetry breaking effects are absorbed in the residual mass which
is reasonably small for these set of lattices. As such, these dynamical fermion configurations are very
valuable and can be used to calculate physical quantities reliably, at least for the mesons. It is suggested
from the study of the nucleon axial coupling  $g_A$ and electromagnetic form factors that the present lattices are
still small and lattices with spatial dimension of 6 fm might be needed in order to control the finite volume errors.

   While combined chiral extrapolation and continuum extrapolation are being carried out with valence domain wall
fermions, we shall explore the viability of employing valence overlap fermions on these DWF configurations. Both
the domain wall fermion and the overlap fermion are chiral fermions. As such, they do not have $O(a)$ errors and
non-perturbative renormalization via chiral Ward identities or the ROM/RI scheme can be implemented relatively
easily~\cite{bcc02,zmd05,gl05}. Furthermore, the overlap fermion has additional desirable features which one can 
take advantage of in order to
improve chiral symmetry as well as the quality of the numerical results. First of all, the numerical implementation
of the overlap fermion allows a precise approximation of the matrix sign function so that the errors on the
sign function and thus the residual mass can be as small as $10^{-10}$ in practice~\cite{cdd04}. The approximation to the exact
chiral symmetry can also be gauged from the Ginsparg-Wilson and the Gell-Mann-Oakes-Renner relations. Its multi-mass
algorithm permits calculation of multiple quark propagators covering
the range from very light quarks to the charm on these sets of DWF lattices. This makes it possible to include
the charm quark for calculations of charmonium and charmed-light mesons using the same fermion formulation for the charm and
light quarks~\cite{dl09}. It is also possible to incorporate partially quenched data in the chiral extrapolation. Since the overlap
operator is a normal matrix, it is easier to calculate its eigenmodes and implement low-mode deflation in the matrix inversion. 
As we shall see, this can speed up the inversion of small quark mass by more than an order of
magnitude with no critical slowing down. Furthermore, these low frequency modes can be used together with the noise
approximation of the high-frequency modes to construct all-to-all or many-to-all correlators. We shall show
that using $Z_3$ grid source on a time slice is quite efficient in reducing variance for both the meson
and baryon correlators. We should point out that although both the overlap and domain wall fermions are chiral
fermions, using overlap valence on DWF gauge configurations with HYP smearing at finite lattice spacing constitutes
a mixed action approach. Mixed action approaches have been studied by many groups such as DWF valence
on staggered fermion sea~\cite{LHPC06}, overlap valence on DWF sea~\cite{amt06}, overlap valence on clover sea~\cite{dfh07}, 
and overlap valence on twisted
fermion sea~\cite{chj09}. It is shown that the valence chiral
fermion has the advantage that it introduces only one extra low-energy constant $\Delta_{mix}$ in the mass of the pseudoscalar
meson with mixed valence and sea quarks which has the same effect as partial quenching.
The mixed
action partially quenched chiral perturbation theory (MAPQ$\chi$PT) which has been worked out for various physical quantities 
with various combination of fermion actions will be the simplest for the combination of overlap valence and DWF sea.

    This manuscript is organized as follows: The formalism for solving linear equation of the overlap operator with low eigenmode 
deflation and $Z_3$ noise
grid for the many-to-all correlators will be given in Sec. 2. The numerical details on the tuning of the negative mass parameter
$\rho$ in the Wilson kernel of the overlap, the speedup due to HYP smearing and low-mode deflation, and the role of the zero 
mode will be presented in Sec. 3.  In Sec. 4, we shall present the calculation of $\Delta_{mix}$ and results for the meson and 
nucleon correlators with point source, $Z_3$ grid source, and $Z_3$ grid source with low-mode substitution. The efficacy of the 
many-to-all approach will be discussed.  We will finish with a summary in Sec. 5.

\section{Formalism}

 The massless overlap operator~\cite{neu98} is defined as
\begin{equation}
D_{ov}  (\rho) =   1 + \gamma_5 \epsilon (H_W(\rho)),
\end{equation}
where $\epsilon (H_W) = H_W /\sqrt{H_W^2}$ is the matrix sign function and $H_W$ is
taken to be the hermitian Wilson-Dirac operator, i.e. $H_W(\rho) = \gamma_5 D_W(\rho)$.
Here $D_w(\rho)$ is the usual Wilson fermion operator, except
with a negative mass parameter $- \rho = 1/2\kappa -4$ in which
$\kappa_c < \kappa < 0.25$. As will explained later in Sec.~\ref{num}, we will use $\kappa = 0.2$ in our
calculation which corresponds to $\rho = 1.5$.

The massive overlap Dirac operator is defined so that at
the tree-level there is no mass or wavefunction renormalization~\cite{cdd04},
\begin{eqnarray}  \label{massive_ov}
D(m) &=& \rho D_{ov} (\rho) + m\, (1 - \frac{D_{ov} (\rho)}{2}) \nonumber\\
&=& \rho + \frac{m}{2} + (\rho - \frac{m}{2})\, \gamma_5\, \varepsilon (H_W(\rho)).
\end{eqnarray}
  Throughout the paper, we shall use the lattice units for dimensionful quantities, except the lattice spacing $a$ will 
be explicit in figures.

\subsection{Deflation}

    It has been advocated~\cite{wil07} that using deflation with low eigenmodes can speed up
inversion of fermion matrices. It has been applied to the hermitian system~\cite{ehn99,dll00} to
speed up the inner loop inversion of the overlap operator and to non-hermitian~\cite{mw07} and
hermitian system with multiple right-hand sides~\cite{so07}. Low-mode deflation has also been applied
to domain decomposition~\cite{lus07}. In addition to speeding up inversions, substituting exact
low eigenmodes in the noise estimation such as in quark loops~\cite{nel01,vk98} and all-to-all
correlators~\cite{gs04,fjo05,kfh07} has demonstrated that better results for the meson two- and three-point
functions can be obtained with reduced errors.

    The massive overlap Dirac operator in Eq.~(\ref{massive_ov}) has the same eigenvectors
as the massless one, we shall consider the massless Dirac overlap $D_{ov}$. Due to the normality of
$D_{ov}$, i.e. $D_{ov}^{\dagger}D_{ov} = D_{ov}D_{ov}^{\dagger}$ and the Ginsparg-Wilson
relation $\{\gamma_5, D_{ov}\} = D_{ov}\gamma_5 D_{ov}$, the eigenvalues of $D_{ov}$ are on a unit circle
with the center at unity. The real and chiral modes are at 0 and 2. Others on the circle are paired with
conjugate eigenvalues. In other words, if $|i\rangle$ is an eigenvector of $D_{ov}$
\begin{equation}
D_{ov}|i\rangle = \lambda_i |i\rangle,
\end{equation}
then $\gamma_5|i\rangle$ is also an eigenvector with eigenvalue $\lambda_i^*$,
\begin{equation}   \label{conjugate}
D_{ov}\gamma_5|i\rangle = \lambda_i^*\gamma_5 |i\rangle.
\end{equation}
To calculate the eigenmodes of $D_{ov}$, one notes that due to normality and $\gamma_5$
hermiticity $D_{ov}^{\dagger} = \gamma_5 D_{ov} \gamma_5$, $\gamma_5$ commutes with
$D_{ov}D_{ov}^{\dagger}$, i.e. $[D_{ov}D_{ov}^{\dagger}, \gamma_5] =0$. Therefore, one
can use the Arnoldi algorithm to search for eigenmodes of $D_{ov}D_{ov}^{\dagger}$
with real eigenvalues $|\lambda_i|^2$ which are doubly degenerate with opposite chirality. To
obtain the eigenmodes of $D_{ov}$, one can diagonalize the two chiral modes in $D_{ov}$.
This is much easier than searching in the complex plane for the eigenmodes of non-normal fermions.
Since the non-zero modes are conjugate pairs (Eq.~(\ref{conjugate})), we need only to save half of
them, e.g.~those with positive imaginary eigenvalues.
When the eigenmodes are calculated, one can proceed with deflation by solving the high frequency
part of the propagator
\begin{equation}  \label{Xh1}
D(m)\;\; |X_{L,R}^H\rangle = (1 - P_L)|\eta_{L,R}\rangle,
\end{equation}
where $P_L = \sum_{i=1}^{n_0+ 2n_l} |i\rangle\langle i|$ is the projection operator to
filter out the low eigenmodes. $n_0$ is the number of zero modes which
are either all left-handed or all right-handed in each configuration. $n_l$ is the
number of non-zero low-frequency modes which come in conjugate pairs. In solving
Eq.~(\ref{Xh1}), we use the conjugate gradient solver (CGNE) for $D(m)D^{\dagger}(m)$. In this case, 
one can utilize the property $D_{ov}D^{\dagger}_{ov} = D_{ov} + D^{\dagger}_{ov}$ to save a
matrix multiplication in each iteration with the chiral source $|\eta_{L,R}\rangle$~\cite{ehn99,dll00}
and the solution $ |X_{L,R}^H\rangle$ has the same chirality as the source.
In this case, Eq.~(\ref{Xh1}) can be written as
\begin{eqnarray}  \label{Xh2}
D(m)\;\; |X_{L,R}^H\rangle &=& |\eta_{L,R}\rangle - \sum_{i=1}^{n_0+n_l} (|i\rangle\langle i|
+ \gamma_5 |i\rangle\langle i|\gamma_5) |\eta_{L,R}\rangle (1 - \frac{1}{2}\delta_{\lambda_i,0}) \nonumber \\
 &=& |\eta_{L,R}\rangle - \sum_{i=1}^{n_0+n_l} (1 \mp \gamma_5)|i\rangle\langle i|\eta_{L,R}\rangle
(1 - \frac{1}{2}\delta_{\lambda_i,0}),
\end{eqnarray}
where the sum is over the zero modes and the low modes on the upper half of the eigenvalue circle.
Although we do not calculate it this way, the high frequency part of the propagator can be written as
\begin{equation}
|X_{L,R}^H\rangle = D^{-1}(m,\rho)\;|\eta_{L,R}\rangle  -  \sum_{i=1}^{n_0+n_l} \lbrack 
\frac{ |i\rangle\langle i|\eta_{L,R}\rangle}
{\rho \lambda_i + m (1- \frac{\lambda_i}{2})} + \frac{\mp \gamma_5|i\rangle\langle i|\eta_{L,R}\rangle}
{\rho \lambda_i^* + m (1- \frac{\lambda_i^*}{2})}\rbrack (1 - \frac{1}{2}\delta_{\lambda_i,0}).
\end{equation}
The total high frequency part of the propagator will be, in the end
\begin{equation}
|X^H\rangle = |X_L^H\rangle + |X_R^H\rangle,
\end{equation}
given that the source $|\eta\rangle = |\eta_L\rangle + |\eta_R\rangle$.

To accommodate the $SU(3)$ chiral transformation with $\delta \psi = T \gamma_5(1 - 1/2 D_{ov}) \psi$~\cite{lus98}
which leads to chirally covariant flavor octet quark bilinear currents in the form $\overline{\psi}\Gamma T
(1 - \frac{1}{2} D_{ov}) \psi$, it is usually convenient to use the chirally regulated field
$\hat{\psi} = (1 - \frac{1}{2} D_{ov}) \psi$ in lieu of $\psi$ in the interpolation field and the currents.
This turns out to be equivalent to leaving unchanged the unmodified interpolation field and currents and
adopting instead the effective propagator
\begin{equation}  \label{Deff}
D_{eff}^{-1} \equiv (1 - \frac{D_{ov}}{2}) D^{-1}(m),
\end{equation}
which also serves to filter out the unphysical eigenmode at $\lambda =2\rho$~\cite{ld05}.

Defining $S \equiv D_{eff}^{-1}$ and $S= S_H + S_L$, where $S_H/S_L$ is the high/low frequency part of the effective propagator,
$S_H$ originating from the source $\eta$ can be obtained, after a few steps of derivation, as
\begin{equation}  \label{Sh}
\langle x|S_H|\eta\rangle \equiv \sum_y S_H(x,y)\eta(y) = \langle x|(1 - \frac{D_{ov}}{2})|X^H\rangle  
=(1 + \frac{m}{2\rho -m})\langle x|X^H\rangle - \frac{1}{2\rho -m}\langle x|(1-P_L)|\eta\rangle.
\end{equation}
It is worthwhile pointing out from Eq.~(\ref{Sh}) that once $|X^H\rangle$ is solved, there is no need to explicitly
multiply $D_{ov}$ on $|X^H\rangle$ which involves an inversion of the kernel $H_W^2$ in the Zolotarev approximation
of the matrix sign function. Similarly, the low frequency part of $S$ can be obtained from spectral decomposition
\begin{eqnarray}  \label{Sl}
\langle x|S_L|\eta\rangle &\equiv& \sum_y S_L(x,y) \eta(y) =  \langle x|(1 - \frac{D_{ov}}{2})|X^L\rangle \nonumber \\
  &=& \sum_{i=1}^{n_0+n_l}
\lbrack \frac{(1 - \frac{\lambda_i}{2}) \langle x|i\rangle\langle i|\eta\rangle}
{\rho \lambda_i + m (1- \frac{\lambda_i}{2})} + \frac{ (1 - \frac{\lambda_i^*}{2})\langle x|\gamma_5|i\rangle
\langle i|\gamma_5|\eta\rangle}{\rho \lambda_i^* + m (1- \frac{\lambda_i^*}{2})}\rbrack
(1 - \frac{1}{2}\delta_{\lambda_i,0}),
\end{eqnarray}
Since the eigenmodes are available for the low frequency modes, one can obtain the all-to-all propagator for
this part of the spectrum
\begin{equation}  \label{Sl_all}
\tilde{S}_L(x, y) =  \sum_{i=1}^{n_0+n_l}
\lbrack \frac{(1 - \frac{\lambda_i}{2}) \langle x|i\rangle\langle i|y\rangle}
{\rho \lambda_i + m (1- \frac{\lambda_i}{2})} + \frac{ (1 - \frac{\lambda_i^*}{2})\langle x|\gamma_5|i\rangle
\langle i|\gamma_5|y\rangle}{\rho \lambda_i^* + m (1- \frac{\lambda_i^*}{2})}\rbrack (1 - \frac{1}{2}\delta_{\lambda_i,0}).
\end{equation}
In the above expressions, we have suppressed the Dirac and color indices.

\subsection{Low-mode substitution}

    It is shown~\cite{gs04,fjo05,kfh07} that when noise is used to estimate the meson two- and three-point
correlation functions in the connected insertion, substituting the noise estimated low-mode part of the correlator
with the exact one improves statistics. Consider, for example, the meson correlator from the local
interpolation fields $O_i=\overline{\psi}\Gamma_i\psi$ where the two-point correlator is
\begin{equation}   \label{meson_corr}
C(t, \vec{p}; \eta) 
 = \sum_{\vec{x},\vec{y}} e^{-i \vec{p}.\vec{x}} \langle {\rm Tr} [\Gamma_1 S(\vec{x},t;\vec{y},0)\eta(\vec{y})\Gamma_2 \gamma_5
S^{\dagger}(\vec{x},t; \vec{y},0)\eta^{\dagger}(\vec{y})\gamma_5]\rangle,
\end{equation}
where the source $\eta$ is on the $t=0$ time slice with support on $\{\vec{y}\}$. Since the quark propagator $S$ is composed of
the low-frequency and high-frequency parts $S = S_H + S_L$, the two-point correlation function can be decomposed
into the following
\begin{equation}
C = C_{HH} + C_{HL} + C_{LH} + C_{LL},
\end{equation}
where
\begin{equation}
C_{LL} = \sum_{\vec{x},\vec{y},\vec{y'}} e^{-i \vec{p}.\vec{x}} {\rm Tr} \langle\Gamma_1 S_L(\vec{x},t;\vec{y},0)
\eta(\vec{y}) \Gamma_2 \gamma_5 S_L^{\dagger}(\vec{x},t; \vec{y'},0)\eta^{\dagger}(\vec{y'})\gamma_5\rangle.
\end{equation}
The noise $\eta$ has the property
\begin{equation}  \label{noise_1}
\langle \eta(\vec{x}) \eta^{\dagger}(\vec{y})\rangle = \delta_{x,y}, 
\end{equation}
where $\langle ... \rangle$ is the noise average.
The standard error due to the noise estimation of the
correlation function averaged over the gauge configurations is given
by~\cite{wl98,dsd09}
\begin{equation}  \label{standard_error}
\epsilon=\sqrt{\frac{\sigma^2_g}{N_g} + \frac{\sigma^2_n}{N_n N_g}}\,\, .
\end{equation}
where $\sigma^2_g/\sigma^2_n$ is the variance of the gauge/noise ensemble, and $N_g/N_n$ is the number of gauge/noise
configurations. How good the approximation is depends on the noise estimation.
Through the numerical study of the quark loop for the energy-momentum tensor with the Wilson
fermion on quenched gauge configurations, it is learned~\cite{dsd09} that $\sigma_n$ is much larger than $\sigma_g$
with $\sigma_n/\sigma_g \sim 27(46)$ for the case with (without) 4-term unbiased subtraction. As such, it
is desirable to replace the noise estimate of $C_{LL}$ with the exact correlator to reduce variance due to the noise 
estimation. The low-mode substituted meson correlator is then
\begin{equation}  \label{C_LL}
C(t, \vec{p})_{sub} = C - C_{LL} + \tilde{C}_{LL},
\end{equation}
with
\begin{equation}
\tilde{C}_{LL} = \sum_{\vec{x},\vec{y}\epsilon G} e^{-i \vec{p}.(\vec{x}-\vec{y})} {\rm Tr} \langle\Gamma_1
\tilde{S}_L(\vec{x},t;\vec{y},0)
\Gamma_2 \gamma_5 \tilde{S}_L^{\dagger}(\vec{x},t; \vec{y},0)\gamma_5\rangle,
\end{equation}
where the sum of $\vec{y}$ runs over the set G with the same support on $\{\vec{y}\}$ as the noise $\eta$.

   In the case of baryon, one can use the $Z_3 (e^{i 2\pi k/3}, k= 0,1,2)$ noise on a time slice
for the quark source due to the property
\begin{equation}   \label{noise_2}
\langle \eta(\vec{x}) \eta(\vec{y}) \eta(\vec{z})\rangle_n = \delta_{x,y} \delta_{y,z},
\end{equation}
so that it is an approximation for the superposition of multiple baryon source with three quarks in each of the baryon
originating from the same spatial location on the support of the $Z_3$ noise.

    Similar to the meson case, one can substitute $C_{LLL}$, which is the part with all three quarks estimated by
$S_L\eta$, with $\tilde{C}_{LLL}$ where all three quark propagators are given in terms of $\tilde{S}_L$. In addition, one
can replace the $C_{HLL}$ part, where one of the quark propagators is $S_H \eta$ and the other two are $S_L\eta$,
by $\tilde{C}_{HLL}$ in which the product of the two $S_L\eta$ is replaced by $\sum_{\vec{y}} S_L(\vec{x},t;\vec{y},0) 
S_L(\vec{x'},t; \vec{y},0)\eta^{\dagger}(\vec{y})$.
\begin{equation}  \label{baryon_sub}
C(t, \vec{p})_{sub} = C - C_{LLL} + \tilde{C}_{LLL} - P\{C_{HLL}\} + P\{\tilde{C}_{HLL}\}
\end{equation}
where $P\{\}$ refers to the set of correlators with permutation of $S_H\eta$ and $S_L\eta$ (or $\tilde{S}_L$)
for the three different quarks in the baryon. It is worthwhile pointing out that $\tilde{C}_{HLL}$ is like $C_{HL}$ 
in the meson in the sense that the error due the noise is from $\sigma_n$ associated with Eq.~(\ref{noise_1}). To the extent that
the baryon correlator is dominated by $C_{LLL}$ and $P\{C_{HLL}\}$ in the time window where the ground state
baryon emerges, the variance reduction with the substitution in Eq.~(\ref{baryon_sub}) is expected to be
similar to that of the meson case.

\section{Numerical Details}  \label{num}

\begin{figure}[ht]
  \centering
  \subfigure[] 
     {\label{1632a}
     {\includegraphics[width=6.5cm,height=4.5cm]{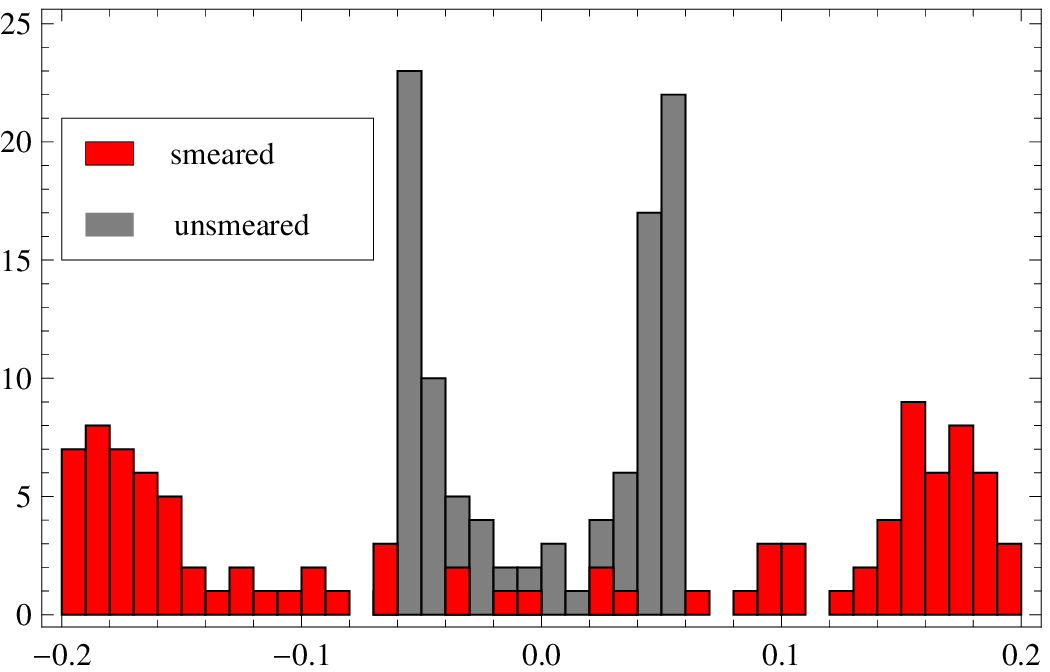}\ \ \ \ }}
  \hspace{0.6cm}
  \subfigure []
     {\label{1632b}
     {\includegraphics[width=6.5cm,height=4.5cm]{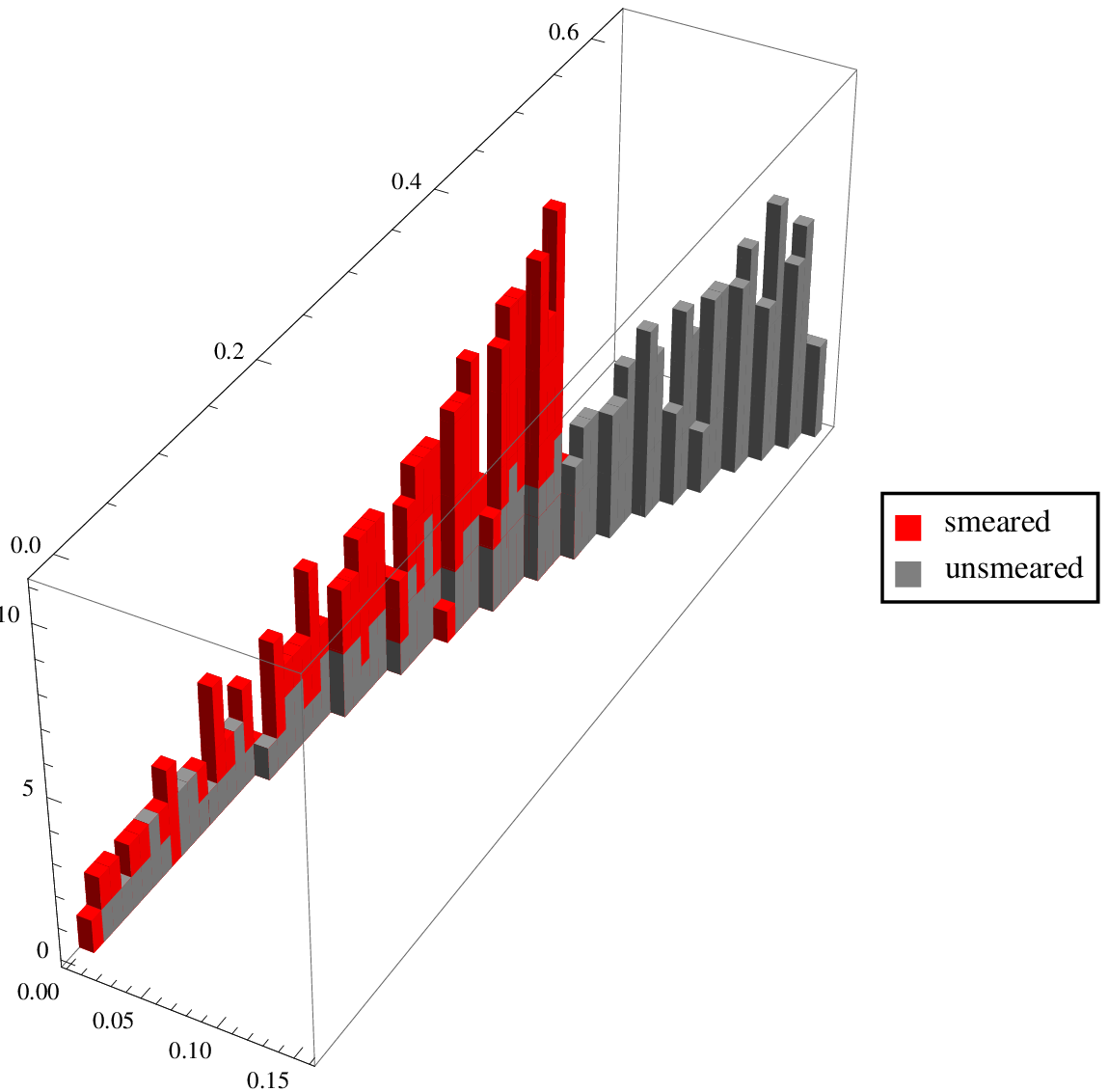}\ \ \ \ }}
  \caption{(color online) (a) The spectra of the lowest 100 eigenvalues for the kernel in the
           inner loop of the overlap fermion for a $16^3 \times 32$ configuration with
           $m_l =0.01 $. (b) The same as (a) for the lowest 200 eigenvalues
           of the outer loop overlap fermion. The unsmeared spectra are
           colored in grey and the HYP smeared spectra are colored in red.}
  \label{1632}
\end{figure}

\begin{figure}[ht]
  \centering
  \subfigure[] 
     {\label{2464a}
     {\includegraphics[width=6.5cm,height=4.5cm]{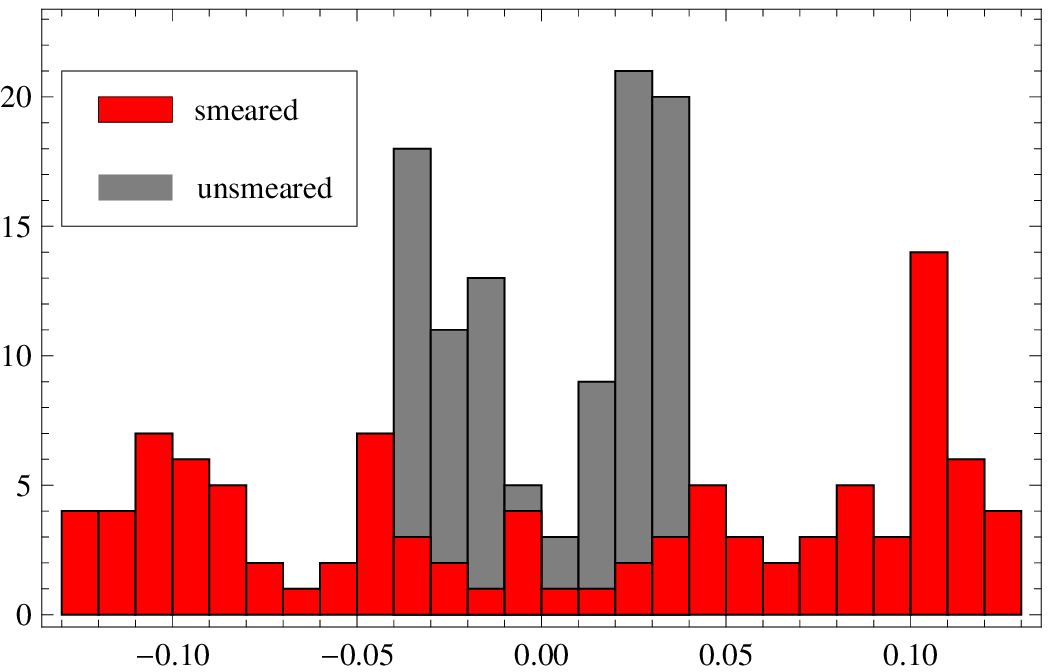}\ \ \ \ }}
  \hspace{0.6cm}
  \subfigure []
     {\label{2464b}
     {\includegraphics[width=6.5cm,height=4.5cm]{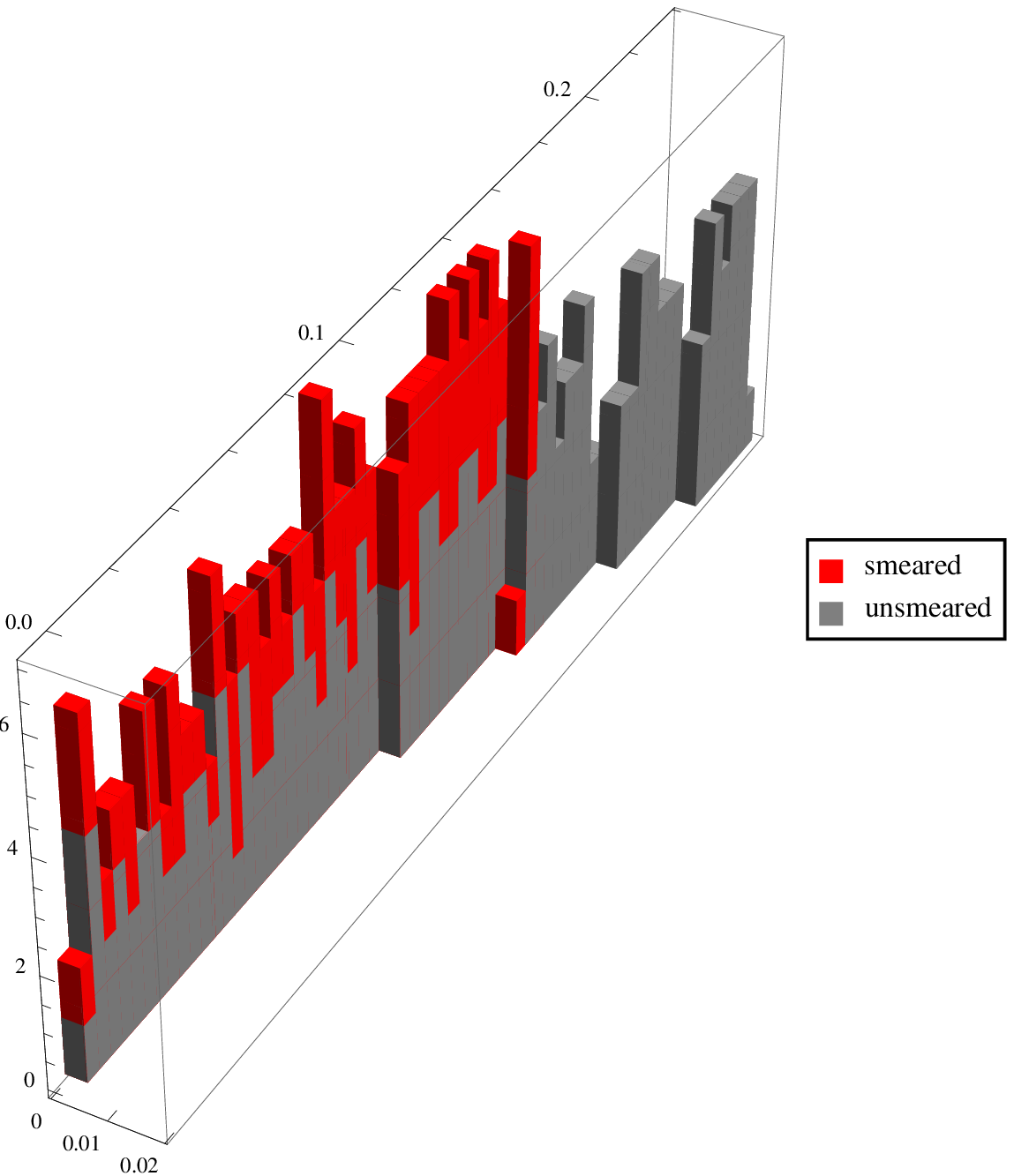}\ \ \ \ }}
  \caption{(color online) (a) The spectra of the lowest 400 eigenvalues for the kernel in the
           inner loop of the overlap fermion for a $24^3 \times 64$ configuration with
           $m_l = 0.005$. (b) The same as (a) for the lowest 200 eigenvalues
           of the outer loop overlap fermion. The unsmeared spectra are
           colored in grey and the HYP smeared spectra are colored in red.}
  \label{2464}
\end{figure}

\begin{figure}[hbt]
  \centering
  \subfigure[] 
     {\label{3264a}
     {\includegraphics[width=6.5cm,height=4.5cm]{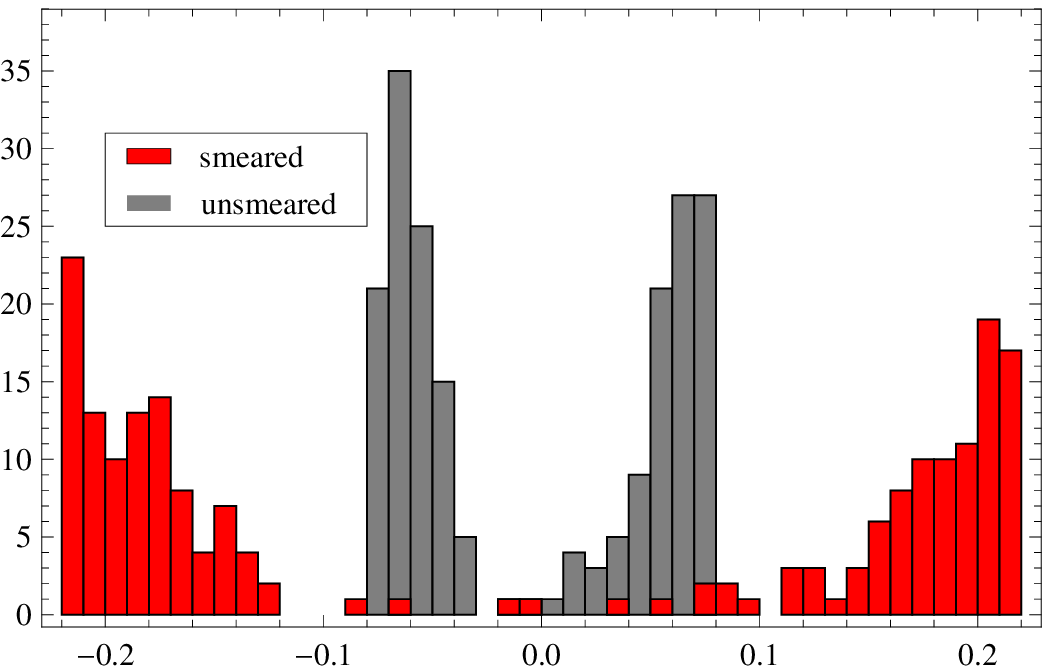}\ \ \ \ }}
  \hspace{0.6cm}
  \subfigure []
     {\label{3264b}
     {\includegraphics[width=6.5cm,height=4.5cm]{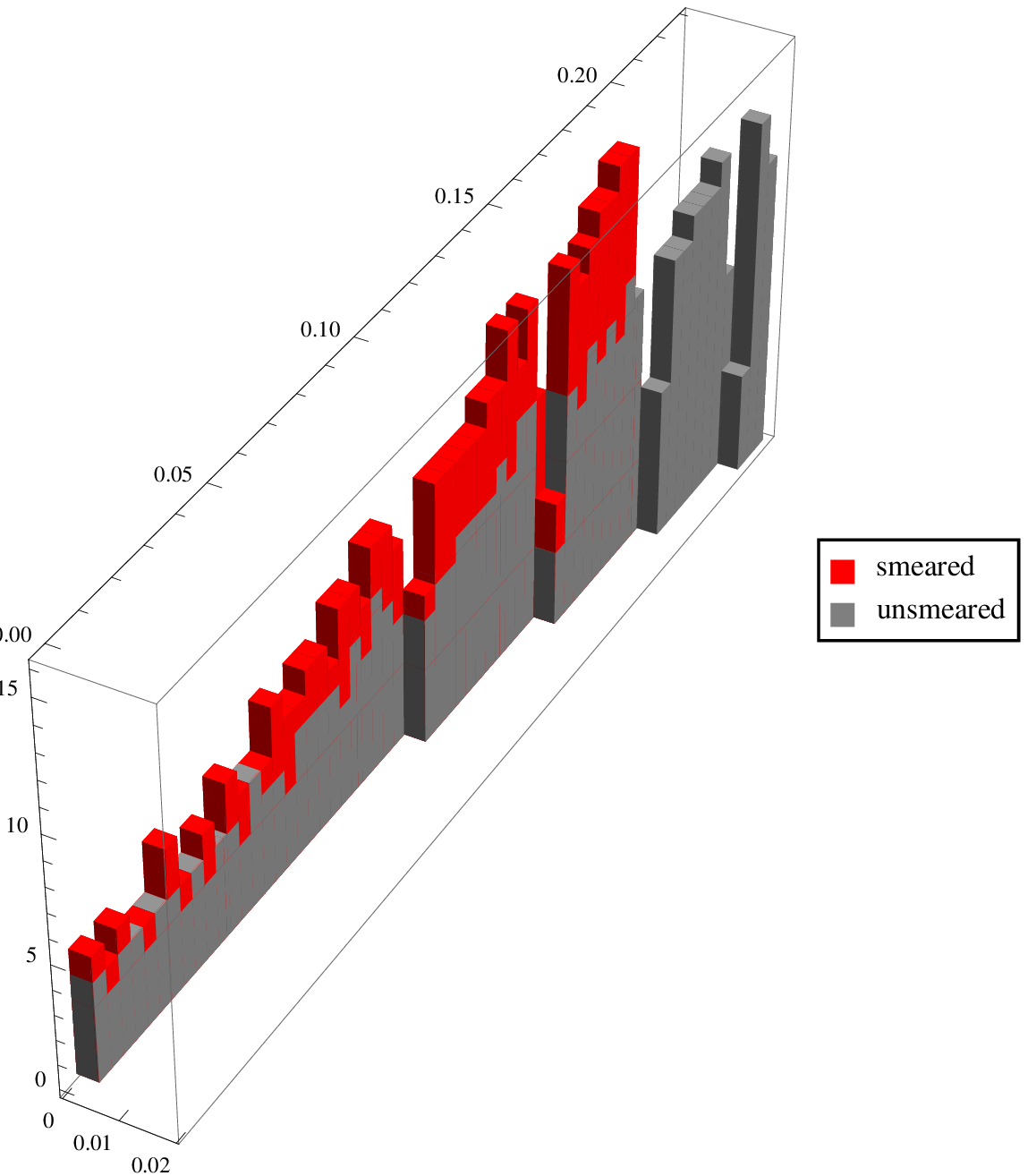}\ \ \ \ }}
  \caption{(color online) (a) The spectra of the lowest 200 eigenvalues for kernel in the
           inner loop of the overlap fermion for a $32^3 \times 64$ configuration with
           $m_l = 0.004$. (b) The same as (a) for the lowest 400 eigenvalues
           of the outer loop overlap fermion. The unsmeared spectra are
           colored in grey and the HYP smeared spectra are colored in red.}
  \label{3264}
\end{figure}

       The overlap propagators are calculated on three sets of lattices of the $2+1$ flavor
domain wall fermion gauge configurations with the four-dimensional sizes of $16^3 \times 32, 24^3 \times 64$
($a^{-1} = 1.73(3)$ GeV), and $32^3 \times 64$ ($a^{-1} = 2.32(3)$ GeV) with several sea quark masses
each. These are generated by the RIKEN-Brookhaven-Columbia (RBC) collaboration and the UKQCD
collaboration~\cite{RBC-UKQCD07,RBC-UKQCD08,maw09}. The matrix sign function in the overlap Dirac operator is
calculated with 14th degree Zolotarev rational polynomial approximation~\cite{vfl02,cdd04}.
For the window [0.031, 2.5], the approximation to the sign function is better than
$3.3 \times 10^{-10}$~\cite{cdd04}. This is sufficiently accurate as the low-mode deflation is used
in the inversion of $H_W$ with HYP smearing in the Zolotarev approximation, which is the inner loop of
the inversion of the overlap operator, and the largest absolute values of the low mode eigenvalues are 
0.2, 0.125, and 0.22 on $16^3 \times 32, 24^3 \times 64$, and $32^3 \times 64$ lattices with 100, 400, and 200 eigenvectors,
respectively. As shown in Figs.~\ref{1632a}, \ref{2464a}, and \ref{3264a}, the largest absolute values of the
projected eigenvalues on the HYP smeared configurations are larger than 0.031, the threshold for high accuracy of 
the approximation of the sign function.

\subsection{Speedup of propagator calculation}

        We employ HYP smearing~\cite{hk01} on the gauge links which has the effect of depleting the
density of the lowest eigenvalues in $H_W$~\cite{dhw05}. As a result, the lowest eigenvalue with HYP smearing 
after deflation with
100 to 200 eigenmodes is about 3 times larger than those without HYP smearing. This leads to $\sim 3$ times
speedup in the number of inner loop CG iterations as was found in a previous study~\cite{dhw05}. This is tabulated in Table~\ref{comparison}. In addition, for the three lattices under study, the numbers of $\rm{H_W}/D_{ov}$ eigenmodes 
used for deflation in the inner/outer loop, the numbers of inner and outer iterations for the cases without deflation, 
with deflation, and with both deflation and HYP smearing. For the comparison study, we used light sea mass at
$m_l = 0.01, 0.005$ and 0.004 for the $16^3 \times 32\, (\rm{lattice 16}), 24^3 \times 64\, (\rm{lattice 24})$
and $32^3 \times 64\, (\rm{lattice 32})$ lattices which are respectively the lowest light sea mass in these three
lattice sets. For the valence quark, we used the quark mass which corresponds to the pion mass at
$\sim 200$ MeV in all three cases. We see from Table~\ref{comparison} that the inner loop iteration number
is reduced by a factor of $\sim 3$ due to HYP smearing. One can see from Figs.~\ref{1632a},~\ref{2464a}, and~\ref{3264a}
that this is due to the fact that after projecting out the small eigenvalues of $H_W^2$ the resultant
lowest eigenvalue with smearing is about a factor of 3 larger than those without smearing. On the other hand,
the number of outer iterations are greatly reduced due to deflation of the low $D_{ov}$ eignemodes. It is interesting
to note that the number of outer iteration for deflation with the smearing case is $\sim 18\% - 25\%$ higher than the
corresponding case without smearing. This is due to the fact that after HYP smearing, the imaginary part of the
highest eigenvalues of the deflated eigenmode $\lambda_{max}$ at $0.0707 \pm i\, 0.434$ (lattice 16),
$0.00857 \pm i\, 0.153$ (lattice 24) and $0.0115 \pm i\, 0.186$ (lattice 32) are 46\%, 51\%, and 22\% smaller than those
of the corresponding eigenvalues without smearing. This can be seen in Figs.~\ref{1632b}, \ref{2464b}, and \ref{3264b}.
In the end, when the total number of iterations are compared, the
speedup with deflation
and smearing to the cases without them are 23, 51, and 79 times respectively for the three lattices for one test
configuration each. The three selected configurations have zero modes. We have also tested configurations
without zero modes. It turns out the total numbers of iterations for both the cases with and without
smearing and deflation are about the same as those of configurations with zero modes. This is so because
the absolute eigenvalues of the lowest eigenmodes for configurations without zero modes (which are of the order $10^{-2}$ for
the $16^3 \times 32$ lattice, $10^{-3}$ for the $24^3 \times 64$ lattice, and  $10^{-4}$ for the $32^3 \times 64$ lattice) 
are comparable to the smallest quark masses on these lattices.

\begin{table}[ht]
\caption{Speedup comparison of inversion with HYP smearing (S)  and deflation (D) of the outer loop. 
The inner and outer iteration numbers are the average of one column in one propagator with 12 columns of color-spin. 
The speedup refers to that between the case of $D+S$ vs the one with neither $D$ nor $S$. The overhead of producing eigenmodes is measured in terms of the
propagators with $D+S$ calculation.
\label{comparison}}
{\footnotesize
\begin{center}
\begin{tabular}{|l|c|ccc|ccc|ccc|}
  \hline
  \multicolumn{1}{|c}{} &\multicolumn{4}{c|}{$16^3\times 32$}
  & \multicolumn{3}{c|} {$24^3 \times 64$} & \multicolumn{3}{c|} {$32^3 \times 64$}\\
  \hline \hline
     & residual & w/o D & D & D+S  & w/o D & D & D+S & w/o D & D & D+S \\
  \hline
  $\rm{H_W}$ eigenmodes & $10^{-14}$ & 100 & 100 & 100 &  400 & 400 & 400 & 200 & 200 & 200 \\
  $D_{ov}$ eigenmodes & $10^{-8}$ & 0 & 200 & 200 &  0 & 200 & 200 & 0 & 400 & 400 \\
  Inner iteration & $10^{-11}$ & 340 & 321 & 108  & 344 & 341 & 107 &  309 & 281 & 101\\
  Outer iteration & $10^{-8}$ & 627 & 72 & 85  & 2931 & 147  & 184 & 4028 & 132 & 156  \\
  \hline
  Speedup    &     &    &  &  23 &     &     & 51 &      &  & 79    \\
  Overhead   &     &    &  \multicolumn{2}{r|} {4.5  propagators}  &  & \multicolumn{2}{r|} {4.9 propagators} &  & 
\multicolumn{2}{r|} {7.9 propagators} \\
  \hline
\end{tabular}
\end{center}
}
 \end{table}

We should point out that the absolute values of the above-mentioned
$\lambda_{max}$ of the low-frequency modes on these lattices are much larger than the small valence quark masses 
(ranging from 0.0014 to 0.01, say) so that the small valence quark does not affect the speed of inversion and, thus,
there is no critical slowing down for the light valence masses in inversions with low-mode deflation. 
Also listed is the overhead for producing eigenmodes of the overlap fermion for deflation. The cost is in the range of
4.5 to 7.9 propagators with both deflation and HYP smearing ($D+S$). This cost is to be amortized when 
more propagators are needed in calculations such as three-point functions and quark loops. To compare with
inversion of the Wilson-type fermion, we timed the inversion of the clover fermion on 10 $2+1$ flavor dynamical clover 
configurations with a size of $32^3 \times 64$ and pion mass of 156 MeV at lattice spacing $a = 0.09$ fm from the 
PACS-CS Collaboration~\cite{aii09}.
Using the CG solver with odd/even preconditioning and no HYP smearing, we find the average inversion of one propagator with 
the same residual of $10^{-8}$ takes $\sim 11892$ iterations. Taking the the product of the inner and outer loop iterations 
for the case of lattice 32 in Table~\ref{comparison}, the overlap inversion with smearing and deflation has about $32\%$ more 
iterations than that of the clover fermion in this case.

\subsection{Tuning of $\rho$}

\begin{figure}[hb]
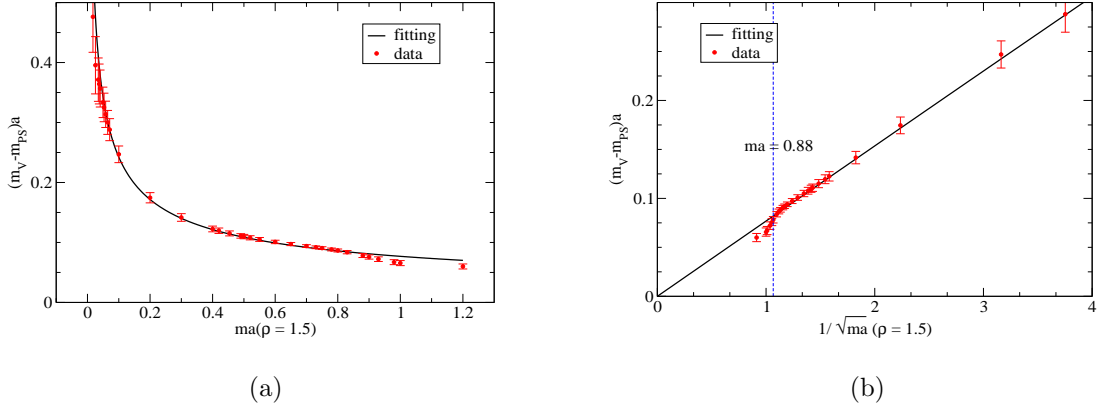

  \centering
  \subfigure[] 
     {\label{hyp15a}
     {\includegraphics[width=6.5cm,height=4.5cm]{figures/hyp_rho15_1.eps}\ \ \ \ }}
  \hspace{0.6cm}
  \subfigure []
     {\label{hyp15b}
     {\includegraphics[width=6.5cm,height=4.5cm]{figures/hyp_rho15_2.eps}\ \ \ \ }}
  \caption{The hyperfine splitting is plotted as a function of $ma$ in (a) and
       $1/\sqrt{ma}$ in (b). This is for $\rho=1.5$.}
  \label{hyp15}
\end{figure}

\begin{figure}[ht]
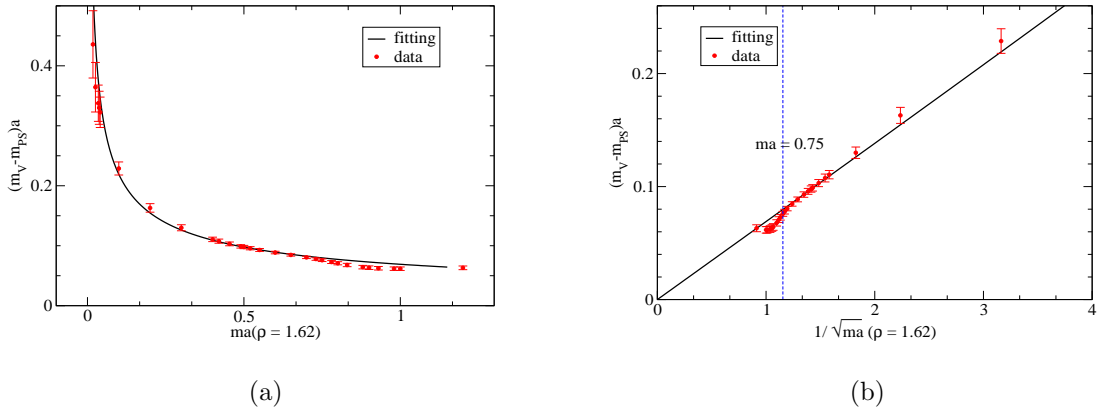

  \centering
  \subfigure[] 
     {\label{hyp162a}
     {\includegraphics[width=6.5cm,height=4.5cm]{figures/hyp_rho162_1.eps}\ \ \ \ }}
  \hspace{0.6cm}
  \subfigure []
     {\label{hyp162b}
     {\includegraphics[width=6.5cm,height=4.5cm]{figures/hyp_rho162_2.eps}\ \ \ \ }}
  \caption{The hyperfine splitting is plotted as a function of $ma$ in (a) and
       $1/\sqrt{ma}$ in (b). This is for $\rho=1.62$.}
  \label{hyp162}
\end{figure}

     We carry out the valence quark propagator calculation with 30 quark masses which cover the range
from the physical pion mass to the charm mass. There will be a concern about the large finite volume effect for
the pion mass as low as the physical one, we shall use those below $m_{\pi} = 200$ MeV for the finite volume study
not in chiral extrapolation. To include the charm mass entails making sure that the heavy mass
will have small enough $O(m^2a^2)$ error to warrant reliable calculation for the charmonium and
charm-light mesons. To this end, we fine-tune the negative mass parameter $\rho$ in Eq.~(\ref{massive_ov})
in the range $1 < \rho < 2$ to minimize the $O(m^2a^2)$ error. We conducted a test on 10 $16^3 \times 32$
($m_l = 0.01$) configurations for the range of $1.059 <\rho < 1.917$ to check the speed of inversion and
 $O(m^2a^2)$ error assessed with the hyperfine splitting (the difference between the vector and pseudoscalar
meson masses). It turns out that $\rho = 1.5$ is close to the optimal choice. It has about the fastest inversion and its
$m^2a^2$ error as measured by the hyperfine splitting in the charmonium is the smallest. To illustrate what one
means to have the smallest $O(m^2a^2)$ error, we plot the hyperfine splitting as a function of $ma$ for the
case of $\rho = 1.5$ and 1.62 in Figs.~\ref{hyp15} and \ref{hyp162} for comparison. The hyperfine splittings for
$\rho = 1.5$ and 1.62 are plotted in Figs.~\ref{hyp15a} and \ref{hyp162a} as a function of $m$. In view of the
fact that the excitation scales for the charmonium and the upsilon as measured from the 2S to 1S and the average
$1{}^3P$ to $1{}^3S$ splittings are about the same, it is argued~\cite{lw78} based on non-relativistic Schr\"{o}dinger
equation that the size of the heavy quarkonium should scale as
\begin{equation}
r_{Q\overline{Q}} \propto \frac{1}{\sqrt{m}}.
\end{equation}

\vspace*{1.5cm}

This prediction is checked against the leptonic decay widths, the fine and hyperfine splittings~\cite{lw78} of
charmonium and upsilon. In the case of the hyperfine splitting, the perturbative spin-spin interaction has
the form $\frac{4\pi\alpha_s}{9}\frac{\sigma_1 \cdot \sigma_2}{m_1 m_2} |\Psi(0)|^2$, where $\Psi(0)$ is the
vector meson wavefunction at the origin which is proportional to $r_{Q\overline{Q}}^{-3/2}$. Thus, the hyperfine
splitting of the heavy quarkonium is expected to scale like
\begin{equation}
\Delta E_{HFS} \propto \frac{1}{\sqrt{m}}.
\end{equation}
to leading order in $m$. Based on this observation, we plot the HFS for $\rho = 1.5$ and 1.62 in
Figs.~\ref{hyp15b} and \ref{hyp162b} in terms of $1/\sqrt{m}$. We see that in both cases, the HFS trends toward
zero as $m$ is heavier than 0.4, except for a few points at the heavy end which show large deviation from the
trend. We interpret this as due to the lattice $O(m^2a^2)$ error. To find a threshold of usable range of $m$
where the estimated $O(m^2a^2)$ error is negligible (or smaller than the statistical error), we fit the HFS to the form
\begin{equation}
\Delta E_{HFS} = \frac{a}{\sqrt{m}}(1 + \frac{b}{m}),
\end{equation}
which includes the next term in large $m$ expansion. From the fit in the range $m = 0.4/0.6$ to 0.8 for
$\rho = 1.5/1.62$ (which correspond to $1/\sqrt{m} = 1.58/1.29$ to 1.12), we find \mbox{$a = 0.0769(6)/0.0690(7)$} and
$b = -0.0002(19)/0.0004(29)$ for $\rho = 1.5/1.62$. We see that the central value of $b/m$ is much smaller than its
error and is thus consistent with zero in both cases.
At the heavy-mass end, the central values of the HFS are outside the fits. We find that at $m = 0.88/0.75$ for 
$\rho = 1.5/1.62$, the central value is beginning to be more
than 2 $\sigma$ away from the fitted HFS curve. We take it to be the critical value beyond
which there is discernible $O(m^2a^2)$ error. This suggests that, given the same relative
deviation, $\rho = 1.5$ has a longer range of
usable $m$ than $\rho = 1.62$. Thus, we decide to adopt $\rho = 1.5$. In this case, the charm quark is
at $\sim 0.73$ where the central value of the HFS is consistent with the fitted curve well within one sigma
(relative error is about 1\%).  
Through study of the $O(m^2a^2)$ error with different lattice spacings for the overlap fermion, it is 
found~\cite{dl07} that the critical mass is insensitive to the lattice spacing and depends mostly on $ma$.
For the $32^3 \times 64$ lattice at $a^{-1} = 2.32$ GeV, the charm mass is at
$m \sim 0.48$ which is much smaller than the critical mass $ma = 0.88$.


We conclude that we can cover the quark mass range from light all the way to the charm 
with the overlap fermion on the three sets of DWF configurations under study. The
critical mass of $ma \sim 0.88$ for a discernible $O(m^2a^2)$ error in the HFS is
higher than that of the quenched case where the $O(m^2a^2)$ error becomes appreciable
(5\%) when $ma \sim 0.5$~\cite{ld05}. This is presumably due to HYP smearing that is
adopted in the present calculation with dynamical fermion configurations. HYP smearing is
also known to improve the locality of the overlap operator~\cite{kov03,dhw05}.

\subsection{Zero mode issue}

\begin{figure}[htbp]
 \vspace*{1cm}
  \centering
   {\includegraphics[width=10cm,height=8cm]{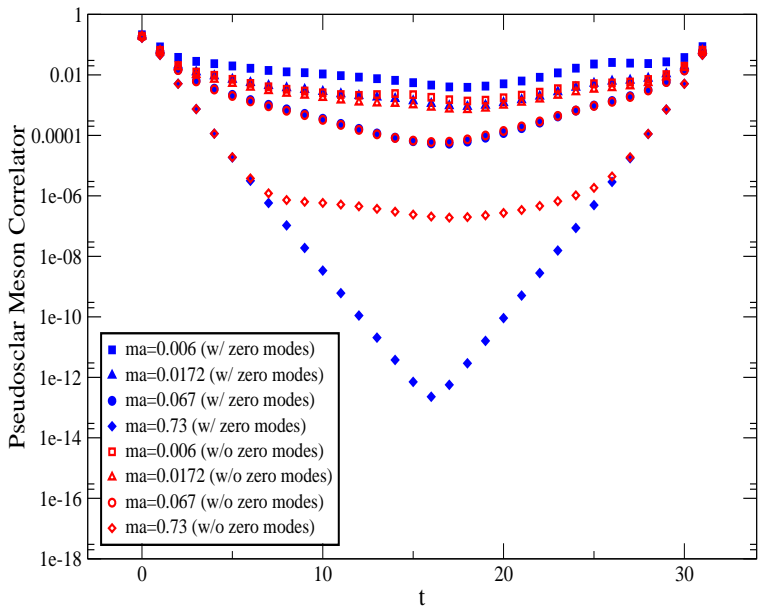}\ \ \ }

 \caption{(color online) The pseudoscalar meson correlators for pion masses at $\sim$ 200, 350, 700 and 2980 MeV which correspond to
the input quark masses at 0.006, 0.0172, 0.067, and 0.73  for a configuration of 
 the $16^3 \times 32$ lattice with and without zero mode contributions.}
  \label{zero_nonzero}
\end{figure}

    The role of zero modes and topology at finite volume has been discussed extensively in the
literature~\cite{ls92}. The calculation of the quark condensate
$\langle \overline{\psi}\psi\rangle$ with the chiral fermion
contains a term from the zero mode contribution $\frac{\langle|Q|\rangle}{mV}$ where $Q$ is the topological
charge of the configuration which, according to Atiya-Singer theorem, equals the difference
between the numbers of left-handed and right-handed zero modes. This term vanishes in the limit
$V \rightarrow \infty$ ($\langle|Q|\rangle$ grows as $\sqrt{V}$) for finite $m$. It is shown~\cite{ls92} that as long
as one is working in the region where $m\Sigma V \gg 1$, the zero mode contribution to the quark condensate
is negligible as the number of zero modes per unit volume goes to zero when the volume approaches infinity.
Based on this and the generalized Gell-Mann-Oakes-Renner relation 
%
\begin{equation}
\frac{1}{V}\int d^4x d^4y \langle \pi^a(x)^{\dagger} \pi^a(y)\rangle = - \frac{2}{m}
\langle\bar{\psi}\psi\rangle,
\end{equation}
where $\pi^a = \bar{\psi}\gamma_5 \tau^a/2 \psi$, it is suggested~\cite{bcc04,ddh02} that,
as long as $m\Sigma V \gg 1$, the contribution of the zeros to the pseudoscalar correlator is negligible. 
In this case, one expects to
obtain the pseudoscalar mass from the exponential fall off of the correlator. To test this
idea, we plot in Fig.~\ref{zero_nonzero} the correlators for the pseudoscalar masses at $\sim 200$, 350, 700 and 2980 MeV.
on the $16^3 \times 32$ lattice for one gauge configuration with and without zero mode contributions
(there are two zero modes in this configuration) for the purpose of illustration. We see that when the 
pion mass is as low as 200 MeV, where $m\Sigma V \sim 1.8$ is not much larger than unity, the pion correlator is 
greatly affected. The pion mass may not change very much, but the spectral weight is reduced by an order of 
magnitude when the zero modes are taken out. When the pion mass is $\sim 350/700$ MeV where  $m\Sigma V \sim 5.5/22$,
we see that the correlators are not affected much when the zero modes are taken out.
This seems to conform with the above idea. However, when we plot the correlator without the
zero modes and examine the heavy quark case where $m\Sigma V \gg 1$ is satisfied, we see that the correlators 
with and without the zero modes differ by several orders of magnitude at large time separation, e.g. $t > 10$. 
This is so because the
zero mode contribution normalized in the present volume is of the order $ > 10^{-8}$ at large time separation;
whereas, the signal for the pseudoscalar meson falls off exponentially with respect to time. Sooner or
later, the signal will fall below the zero mode contribution. In other words, the zero mode at any finite
volume is part of the physical spectrum. Except for quark condensate and other rare cases, one cannot separate out 
the zero mode contribution from the rest of the spectrum for physical observables in general. Even though it may not make 
much of a difference numerically for the meson
correlator at relatively short range of $t$ when $m\Sigma V \gg 1$, the large time separation will sooner or later 
be affected and this problem is more acute for the heavy quark.

\section{Results}

   Since the overlap fermion is calculated on configurations generated with domain wall fermions,
 this constitutes a mixed action approach to chiral fermions. Mixed action approaches have been studied by several groups,
such as DWF valence on staggered fermion sea~\cite{LHPC06}, overlap valence on DWF sea~\cite{amt06}, overlap valence on 
clover sea~\cite{dfh07}, and overlap valence
on twisted fermion sea~\cite{chj09}. In view of the fact that it is numerically intensive to simulate chiral fermions
(DWF or overlap), it is practical to use the cheaper fermion formulation for generating gauge configurations and
the chiral fermion for the valence as an expedient approach toward full unquenched QCD simulation with chiral fermions.
Many current algebra relations depend only on the chiral property of the
valence sector. The mixed action theory with different fermions for the valence and the sea is a generalization of the
partially quenched theory with different sea and valence quark masses.
The mixed action partially quenched chiral perturbation theory (MAPQ$\chi$PT) has been developed for
Ginsparg-Wilson fermion on Wilson sea~\cite{brs03} and staggered sea~\cite{bbr05}, and has been worked out for
many hadronic quantities to next-to-leading order (NLO), such as pseudoscalar masses and decay 
constants~\cite{brs03,bbr05,cow07},
isovector scalar $a_0$ correlator~\cite{gis05,pre06,alv08,wlo09},
heavy-light decay constants~\cite{ab06}, and baryon masses~\cite{tib05,wlo09}.

In the mixed action chiral perturbation theory with chiral valence fermion, it is shown~\cite{brs03} that to
NLO, there is no $\mathcal{O}(a^2)$ correction to the valence-valence meson mass due to
the chiral symmetry of the valence fermion. Furthermore, both the chiral Lagrangian and the chiral extrapolation
formulas for hadron properties to the one-loop level (except $\theta$-dependent quantities) are independent of the
sea fermion formulation~\cite{cwo07}. The LO mixed-action chiral Lagrangian invokes only one more term with
$\mathcal{O}(a^2)$ discretization dependence which is characterized by a low energy constant $\Delta_{mix}$.
The LO pseudoscalar meson masses are given as
\begin{eqnarray}   \label{mixed_mass_relation}
m_{vv'}^2 &=& B_{ov}(m_v +m_{v'}), \nonumber \\
m_{vs}^2 &=& B_{ov}m_v + B_{dw}(m_s + m_{res}) + a^2\Delta_{mix}, \nonumber \\
m_{ss'}^2 &=& B_{dw}(m_s + m_{s'} + 2 m_{res}),
\end{eqnarray}
where $m_{vv'}/m_{ss'}$ is the mass of the pseudoscalar meson made up of valence/sea quark and antiquark.
$m_{vs}$ is the mass of the mixed valence and sea pseudoscalar meson. Up to numerical accuracy, there is no
residual mass for the valence overlap fermion. The DWF sea has a residual mass $m_{res}$ which vanishes as
$L_S \rightarrow \infty$. The $\Delta_{mix}$ enters in the mixed meson mass $m_{vs}$ and is an
$\mathcal{O}(a^2)$ error which vanishes at the continuum limit. We should note that, unlike the partially
quenched case, even when the quark masses in the valence and sea match, the unitarity is still violated due
to the mixed action. The degree of unitarity violation at finite lattice spacing depends on the size of $\Delta_{mix}$.

\subsection{Calculation of $\Delta_{mix}$}

    $\Delta_{mix}$ has been calculated for pseudoscalar mesons for DWF valence and staggered fermion sea~\cite{ow08,alv08} which
gives $\Delta_{mix} \sim (708 {\rm MeV})^4$~\cite{ow08} and $\sim (664 {\rm MeV})^4\pm (437 {\rm MeV})^4$~\cite{alv08}. It is also calculated for overlap valence and clover sea which yields $\Delta_{mix} = (872 {\rm MeV})^4 \pm (693 {\rm MeV})^4$~\cite{dfh07a}. This means that for a valence pion of 300 MeV, the $\Delta_{mix}$ produces, at $a = 0.12$ fm, a shift of $\sim 102 - 251$ MeV for 
these cases which is quite large.

\begin{figure}[ht]
  \centering
  \subfigure[] 
     {\label{wrap1}
     {\includegraphics[width=6cm,height=6cm]{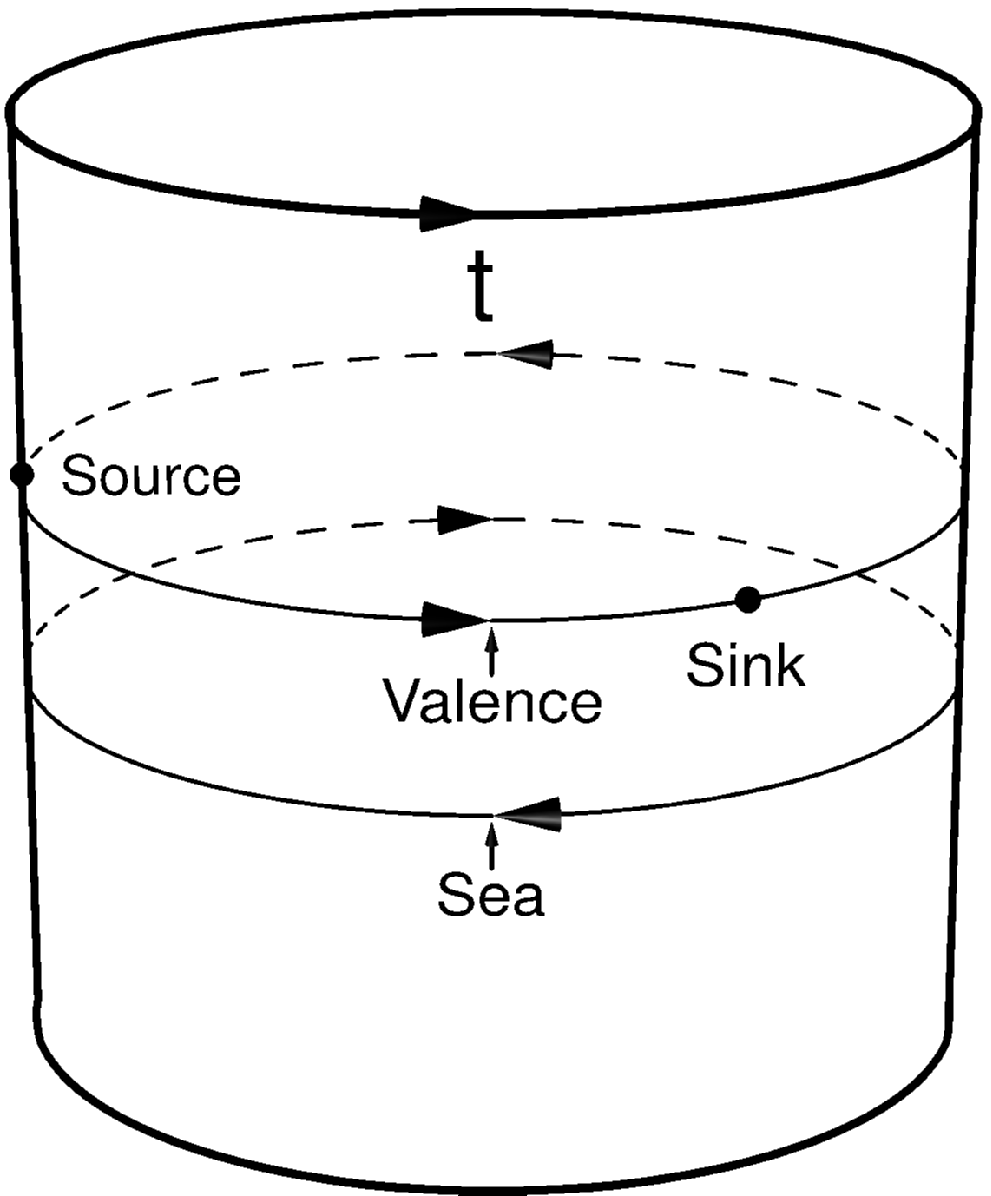}\ \ \ \ }}
  \hspace{0.6cm}
  \subfigure []
     {\label{wrap2}
     {\includegraphics[width=6cm,height=6cm]{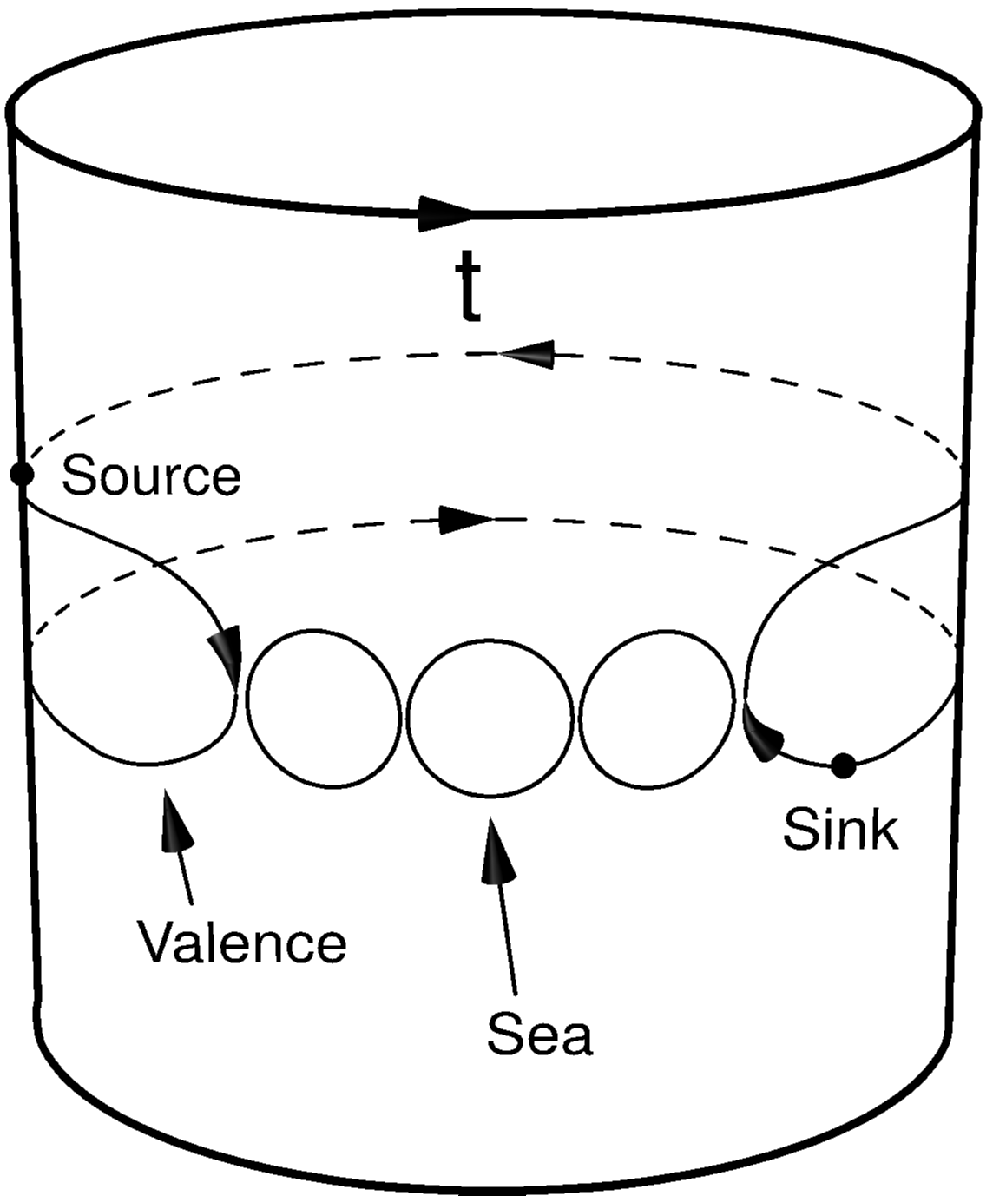}\ \ \ \ }}
  \caption{Cartoon showing the quark lines which form wrap-around $\pi$ and $\eta$ mesons in the $a_0$
  correlator. (a) Both the valence and the sea wrap around the time direction and form two pions which are propagating 
in different time directions. (b) The annihilation diagram where only the valence wraps around the time boundary. 
Together with (a), it produces $\pi$ and $\eta(\eta')$ propagating in different time directions.}
  \label{wrap}
\end{figure}

Here we shall estimate $\Delta_{mix}$ in our case with overlap valence on DWF sea. To do so, we shall
examine the meson state which wraps around the time boundary. It is known that a two meson interpolation
field can produce meson states with two mesons propagating along opposite time directions~\cite{dos08,pm09,pdl10}. On the
other hand, the $a_0$ isovector scalar meson interpolation field $\bar{u}d$, together with the
quark loop from the sea, can produce a $\pi$ and $\eta(\eta')$ propagating in
different time direction~\cite{ddl08}. This is illustrated in Fig.~\ref{wrap} where Fig.~\ref{wrap1} displays
the situation with both the valence and the sea quarks wrapping around the time boundary forming two pions propagating
in different time directions; whereas, Fig.~\ref{wrap2} shows the annihilation diagram where the valence quark-antiquark
pair wraps around the time boundary while the sea quark loops do not. Together, Fig.~\ref{wrap1} and  Fig.~\ref{wrap2}
form an opposite-time propagating $\pi$ and $\eta(\eta')$ pair. The $a_0$ correlator has, thus, the following form
\begin{eqnarray}  \label{wrap_correlator}
C_{a_0} &=& \sum_{i} W_i (e^{-E_it} + e^{-E_i(T-t)}) + W_{\pi\eta} (e^{-m_{\pi}t - m_{\eta}(T-t)}
+ e^{-m_{\eta}t - m_{\pi}(T-t)}), \nonumber \\
 &=& \sum_{i} 2 W_i e^{-E_i T/2} \cosh(E_i(T/2-t)) \nonumber \\
 &+& 2 W_{\pi\eta} e^{-(m_{\pi}+m_{\eta})T/2} \cosh((m_{\eta}-m_{\pi})(T/2-t)),
\end{eqnarray}
where $E_i$ is the energy of one- or two-meson state which propagates in the same time direction. Notice that
the second term in Eq.~(\ref{wrap_correlator}), which represents the $\eta$ and $\pi$ wrapping around the time boundary,
has an exponential falloff which is proportional to the mass difference of $\eta$ and $\pi$, i.e $m_{\eta}-m_{\pi}$.
Since this is smaller than all the other states in the $a_0$ correlator~\footnote{We should note that there are 
states where $\pi$ and $\eta$ move back to back with non-zero momenta which have smaller energy differences, but they 
will be suppressed by the corresponding pre-factor $e^{-(E_{\pi} + E_{\eta})T/2}$ as compared to
$e^{-(m_{\pi} + m_{\eta})T/2}$ in Eq.~(\ref{wrap_correlator}) for the the zero-momentum $\pi-\eta$ state for large $T$.}
it appears as the lowest state in
the longest time separation in the correlator. This low-lying state causes problems for fitting the scalar correlator
to obtain the $a_0$ meson~\cite{ddl08}. 

We shall take advantage of the existence of this state due to the finite
time extent to extract $\Delta_{mix}$. As we see in Fig.~\ref{wrap}, the structure of Fig.~\ref{wrap2} is
complicated. Yet, Fig.~\ref{wrap1} is simple in that it involves the mass difference of two pions, not
$\eta$ and $\pi$ as in Eq.~(\ref{wrap_correlator}). For the equal valence quark mass case, the mass difference of the 
two pions vanishes and one does not obtain any
information about $\Delta_{mix}$. But if the two valence quark masses are not the same, the mass difference
of the two pions becomes
\begin{equation}
m_{v_1s} - m_{v_2s} = \sqrt{B_{ov}m_{v_1} + B_{dw}(m_s + m_{res}) + a^2\Delta_{mix}}
- \sqrt{B_{ov}m_{v_2} + B_{dw}(m_s + m_{res}) + a^2\Delta_{mix}},
\end{equation}
where $m_{v_1s}/m_{v_2s}$ is the pion made of quarks with $v_1/v_2$ overlap fermion and the $s$ domain-wall
fermion. In this case, one can extract $\Delta_{mix}$. To do so, one needs to remove the annihilation diagram
in Fig.~\ref{wrap2} and all the states which propagate in the same time direction. This can be achieved by calculating the valence 
propagators with both anti-periodic and
periodic boundary conditions in time (the DWF sea has anti-periodic B.C.) and take the difference of the two correlators.
For the annihilation diagram in Fig.~\ref{wrap2}, the valence quark traverses the time boundary an even number of times so
that it is independent of the time boundary condition. On the contrary, the valence quark in Fig.~\ref{wrap1} traverses the time 
boundary an odd number of times. So, taking the difference between correlators with periodic and anti-periodic time B.C. for
the valence cancels out the annihilation diagram in Fig.~\ref{wrap2} as well as the contribution from states propagating in the 
same time direction and one is left with the contribution in Fig.~\ref{wrap1}.

   As a first attempt to extract $\Delta_{mix}$, we consider the difference of scalar correlators from the $24^3 \times 64$
DWF lattice (light sea mass at $m_l = 0.005$) with periodic and anti-periodic B.C. in time. In this case,
the difference correlator at large time will be given by
\begin{equation} \label{equal_BC}
\Delta C_{a0}= C_{a_0}^{P} - C_{a_0}^{AP} \longrightarrow 4 W_{\pi_1\pi_2} e^{-(m_{v_1s}+m_{v_2s})T/2} 
cosh\{(m_{v_1s}-m_{v_2s})(T/2-t)\}.
\end{equation}

\begin{figure}[ht]
  \centering
   \includegraphics[width=6.5cm,height=4.5cm]{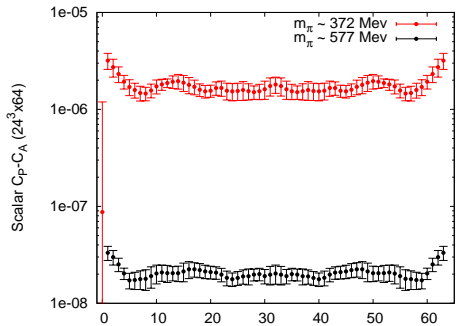}\ \ \ \ 
 \caption{The difference of the scalar ($a_0$) correlators with anti-periodic and periodic time boundary conditions for
 the two equal valence masses which correspond to $m_{\pi} = 372$ and 577 MeV.}
  \label{equal_mass}
\end{figure}

\begin{figure}[hbt]
  \centering
  \subfigure[] 
     {\label{pi2_m}
     {\includegraphics[width=6cm,height=5cm]{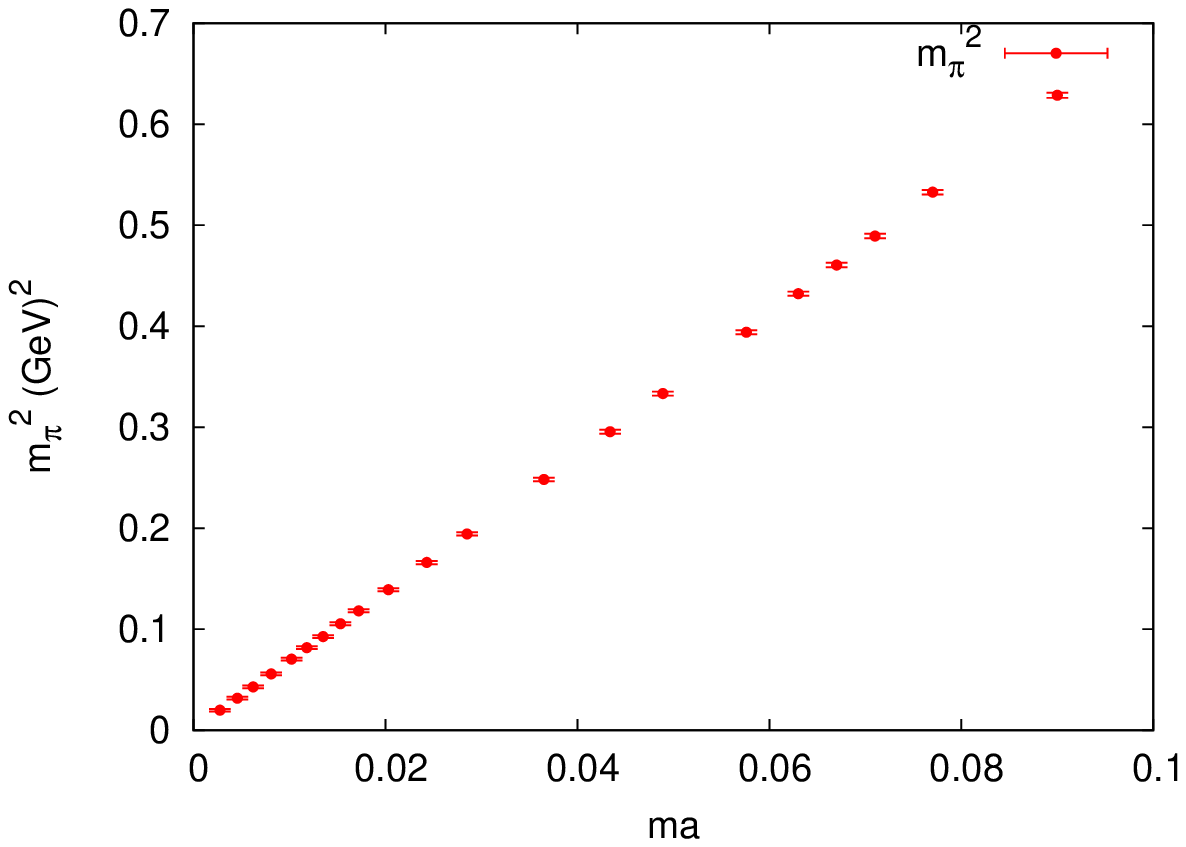}\ \ \ \ }}
  \hspace{0.6cm}
  \subfigure []
     {\label{pi2divm}
     {\includegraphics[width=6cm,height=5cm]{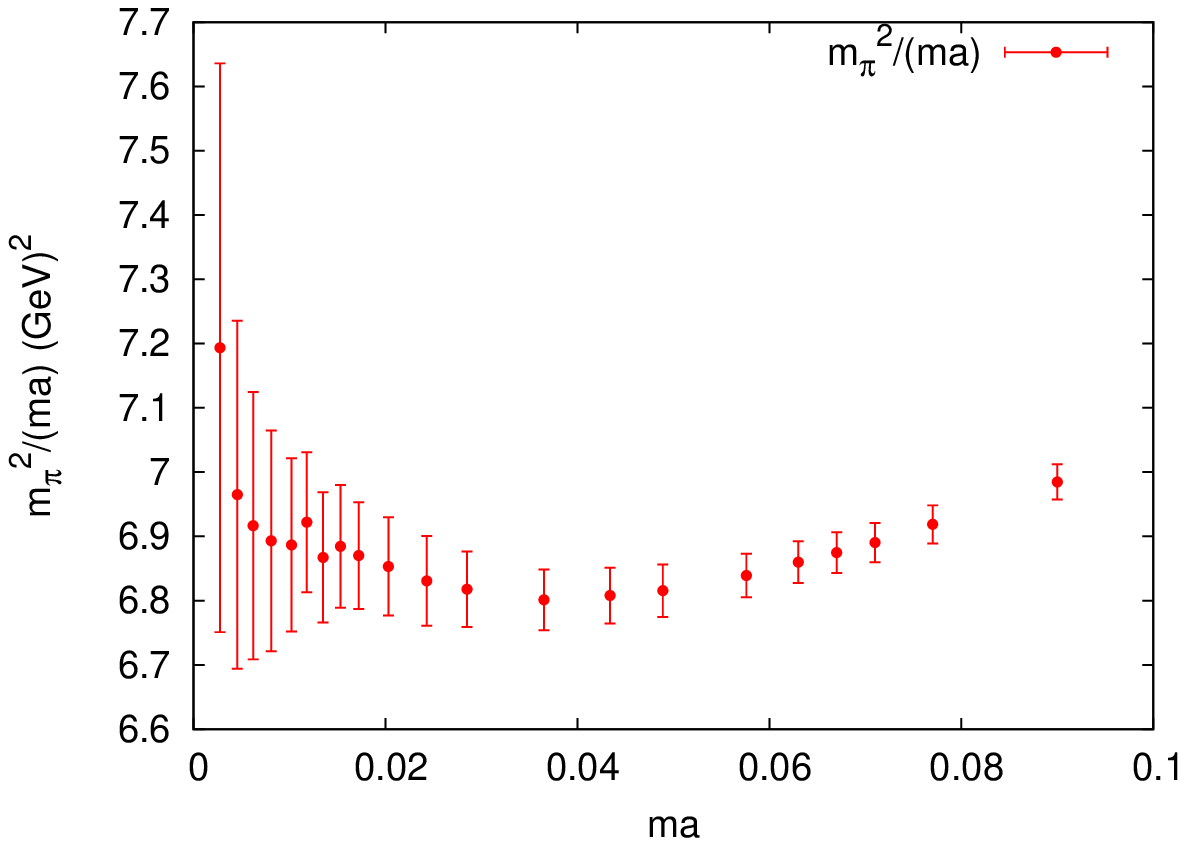}\ \ \ \ }}
  \caption{(a) $m_{\pi}^2$ is plotted as a function of $ma$ for the $24^3 \times 64$ lattice
    with $m = [0.00275, 0.15]$. (b) $m_{\pi}^2/ma ({\rm GeV})^2$ is plotted vs $ma$ for the same range of
    quark masses.}
  \label{mPS2m}
\end{figure}

As a first check, we plot in Fig.~\ref{equal_mass} such a difference correlator for the equal valence case
(i.e. $m_{v_1} = m_{v_2}$). We expect from Eq.~(\ref{equal_BC}) that the correlator should be independent of $t$.
As we see in Fig.~\ref{equal_mass} where such correlators are plotted for $m_{v_1}=m_{v_2} = 0.0203$ and 0.0489
which correspond to pion masses at $\sim 372$ and 577 MeV, the correlators are indeed quite flat. This is consistent
with our expectation that the large $t$ behavior depends on the mass difference $m_{v_1s}-m_{v_2s}$ which is zero
in this case. To extract $\Delta_{mix}$, we want to find a range of quark mass where the tree-level linear mass relation
between $m_{vv}^2$ and $m_v$ holds so that we can use Eq.~(\ref{mixed_mass_relation}). We plot $m_{vv}^2$ and
$m_{vv}^2/m_v$ from the $24^3\times 64$ lattice with $m_l = 0.005$ as a function of $m_v$ in Fig.~\ref{mPS2m}.

 We see that the ratio
$m_{vv}^2a^2/m_va$ in Fig.~\ref{pi2divm} is fairly flat for the range $m_va \sim 0.0203 - 0.0489$. We shall take $m_{v_1}$ and
$m_{v_2}$ from this range and fit the correlators to find $m_{v_1s}- m_{v_2s}$ which can be expressed in terms of
the corresponding pseudoscalar masses and $\Delta_{mix}$,
\begin{equation}   \label{delta_mix}
m_{v_1s}- m_{v_2s} = \sqrt{\frac{1}{2}(m_{v_1v_1}^2 + m_{ss}^2) + a^2\Delta_{mix}} -
\sqrt{\frac{1}{2}(m_{v_2v_2}^2 + m_{ss}^2) + a^2\Delta_{mix}}.
\end{equation} 
>From the separately calculated $m_{v_1v_1}$ and $m_{v_2v_2}$ with the valence overlap fermion and $m_{ss}$ 
calculated with DWF~\cite{RBC-UKQCD08}, we can extract $\Delta_{mix}$.

Since the range of $m_va \sim 0.0203 - 0.0489$ is narrow and $m_{v_1v_1}$ and $m_{v_2v_2}$ are close, the errors on
the extracted $\Delta_{mix}$ from Eq.~(\ref{delta_mix}) are large. We take several combinations of $m_{v_1v_1}$ and $m_{v_2v_2}$
in the range [0.0203, 0.0489] and obtain an average $\Delta_{mix}$ 
\begin{equation}
a^2 \Delta_{mix} = - 0.0112(44)\, {\rm GeV}^2,
\end{equation}
for the $24^3 \times 64$ lattice with 50 configurations at $m_l = 0.005$. With $a^{-1} = 1.73$ GeV, we obtain
\begin{equation}
\Delta_{mix} = - (427  {\rm MeV})^4 \pm (338 {\rm MeV})^4.
\end{equation}

This is quite small. To compare with those from other mixed actions, we notice that the central value is $\sim$ 7 times smaller
than the case of DWF valence on staggered sea~\cite{ow08,alv08} and $\sim$ 18 times smaller than that of overlap
on Wilson sea~\cite{dfh07}. To put the magnitude in perspective, consider a 300 MeV pion on the
$24^3 \times 64/32^3 \times 64$ lattice with $a \sim 0.12/0.085$ fm, the shift in mass due to $\Delta_{mix}$ is
$\sim 19/10$ MeV which is substantially smaller than the $\sim 102 -251$ MeV for the other mixed actions as alluded 
to earlier. As mentioned above, the calculation of $\Delta_{mix}$ using the boundary condition method gives
large errors. At this stage, we are more interested in finding out how large $\Delta_{mix}$ is roughly to see if it is
practically small enough to carry out chiral extrapolation with MAPQ$\chi$PT. For a more precise value, we shall
use the mixed valence DWF and overlap propagators to directly evaluate $\Delta_{mix}$  and check scaling as is done in
Refs.~\cite{ow08,alv08,dfh07}.

Coming back to the correlators in Fig.~\ref{equal_mass}, we notice that the magnitudes of these two correlator differ by 
two orders of magnitude, a feature which is unusual for meson correlators with pion masses which are not that different. In this
case, we note that there is a pre-factor $e^{-(m_{v_1s}+m_{v_2s})T/2}$ in Eq.~(\ref{equal_BC}) which can be sensitive
to slightly different pion mass when $T$ is large. Since we expect the ratio of the spectral weight $W_{\pi_1\pi_2}$ in Eq.~(\ref{equal_BC}) 
for the two correlators with pion masses at $\sim 372$ and 577 MeV to be within $\sim 20\%$ from unity when the mass dependence 
of the matrix element and the normalization factor are taken into account, the primary difference of the correlators in
Fig.~\ref{equal_mass} for the equal masses case should come from the exponential pre-factor. Taking $\Delta_{mix}$ into
account, the ratio of the pre-factor is 88(13) with $T=64$. This turns out to be quite close to the jackknife ratio of the 
calculated correlators in the time range [11,54] which is 83(9). This lends further support for the existence of the wrap-around states.

\subsection{$Z_3$ grid source and low-mode substitution}

Noise has been used to estimate quark loops as well as propagators in the all-to-all
correlators. In particular, $Z_2$ noise has been introduced to estimate the quark loops~\cite{dl94} and it is 
shown~\cite{bmt94,dl94} that its
variance is minimal since, unlike the Gaussian noise, it receives no contribution from the diagonal
matrix elements of the quark propagator. Complex $Z_2$ (or $Z_4$) has been adopted in many quark loop
calculations~\cite{dll95,dll96,dlw98,ddd09} and the stochastic estimation of determinants~\cite{tdl98}.
However, the volume $Z_2$ noise is not a good estimator for connected insertion calculation where the hadron
correlator $C(t,0)$ is needed for large time separation. In this case, the signal falls off exponentially
($C(t,0) \sim e^{-mt}$),
yet the variance decreases only as the inverse power of the noise number~\cite{dl94}. To alleviate this difficulty,
dilution of the noise is suggested so that the noise is applied to one time slice at a time and 
supplemented with low-mode substitution~\cite{fjo05,kfh07}.

   In the following, we shall consider the $Z_3$ noise which can be used for
 baryons as well as mesons.

\subsubsection{Signal-to-noise ratio}

      We should first remark that the noise wall source, by itself, does not reduce errors as compared to the point source.
 To see this, we shall consider the variance of the meson and baryon correlators. Besides the large time behavior
first considered by Lepage~\cite{lep89}, there are pre-factors associated with the noise source. The meson correlator
in Eq.~(\ref{meson_corr}) with a noise wall source has the following behavior at large $t$,
\begin{equation}
C_M(t, \vec{p}=0) \sim V_3 e^{-m_Mt},
\end{equation}
where $V_3$ is the three-volume of the noise with its support on a time slice. This comes from the noise estimate
with $\sum_{\vec{x},\vec{y}} \langle \eta^{\dagger}(\vec{x})\eta(\vec{y})\rangle \propto V_3$. Thus, the signal from the 
noise estimator 
is larger than that of a point source by a factor of $V_3$ according to our normalization convention of the noise. 
On the other hand, the variance of the correlator at large $t$ is
\begin{equation}  \label{var_meson}
N \sigma_M^2 (t) \approx \langle G_M(t)^2\rangle - \langle G_M(t) \rangle^2,
\end{equation}
where $N= N_g \times N_n$ and $G_M(t)$ is the meson propagator in each gauge and noise configuration as defined in terms
of the meson correlator in Eq.~(\ref{meson_corr}), i.e.
\begin{equation}
C(t) = \langle G_M(t)\rangle.
\end{equation}
In the case of the flavor non-singlet meson, the lowest energy state in the variance correlator of Eq.~(\ref{var_meson})
is about the mass of two pions. The noise from the first term contributes a volume squared factor from the four
quark propagator, i.e.
\begin{equation}
\sum_{\vec{x},\vec{y},\vec{x'},\vec{y'}} \langle \eta^{\dagger}(\vec{x})\eta(\vec{y})\eta^{\dagger}(\vec{x'})\eta(\vec{y'})\rangle 
\propto V_3^2.
\end{equation}
Therefore, at large $t$
\begin{equation}
\sigma_M(t) \approx \frac{V_3}{\sqrt{N}} e^{-m_{\pi}t}
\end{equation}
The signal to noise ratio is
\begin{equation}
\frac{C_M(t,\vec{p}=0)}{\sigma_M(t)} \approx \sqrt{N} e^{-(m_M-m_{\pi})t}.
\end{equation}
The volume factor cancels out and there is no gain in statistics with the noise wall source as compared
to a point source.
For the pion correlator, the signal-to-noise ratio is nearly constant at large $t$ as noted before~\cite{lep89}.

    The baryon case is different. Consider the nucleon correlator which has the generic form
\begin{equation}
C_N(t,\vec{p}=0) \sim \langle S(t,0)S(t,0)S(t,0)\rangle,
\end{equation}
where $S(t,0)$ is the $u/d$ quark propagator and it is produced with
a $Z_3$ wall source. We have suppressed the associated $\gamma$ matrices in this expression.
At large $t$,
\begin{equation}
C_N(t,\vec{p}=0) \approx V_3 e^{-m_Nt}.
\end{equation}
The $V_3$ factor comes from the sum of the noises at the source end,
\begin{equation}
\sum_{\vec{x},\vec{y},\vec{z}} \langle \eta(\vec{x})\eta(\vec{y})\eta(\vec{z})\rangle = V_3.
\end{equation}
As for the variance
\begin{equation}
N \sigma_N^2 (t) \approx \langle S^3(t,0) S^{\dagger\, 3}(t,0)\rangle - C_N^2(t,0),
\end{equation}
the lowest mass state in the first term at large $t$ is 3 pions  which is lower than
that of the second term
which falls off like $e^{-2 m_Nt}$. Besides the large time behavior, the variance has a
pre-factor from the noise
\begin{equation}
\sum_{\vec{x},\vec{y},\vec{z},\vec{x'},\vec{y'},\vec{z'}} \langle \eta^{\dagger}(\vec{x})\eta^{\dagger}(\vec{y})\eta^{\dagger}(\vec{z})
\eta(\vec{x'})\eta(\vec{y'})\eta(\vec{z'})\rangle \propto  V_3^3,
\end{equation}
so that the signal-to-noise ratio is
\begin{equation}  \label{s2n_N}
\frac{C_N(t,\vec{p}=0)}{\sigma_N(t)} \approx \sqrt{\frac{N}{V_3}} e^{-(m_N-3/2 m_{\pi})t}.
\end{equation}
It shows that, besides the familiar large time fall off, there is an additional factor
of $V_3^{-1/2}$ due to the noise. This makes the $Z_3$ wall source worse than the
point source statistically.

\begin{figure*}[hbtp]
\begin{center}
\mbox{
	\leavevmode
	\subfigure [ Pion ]
	{ \label{pi_relerr}
	  \includegraphics[width=5cm,height=6cm]{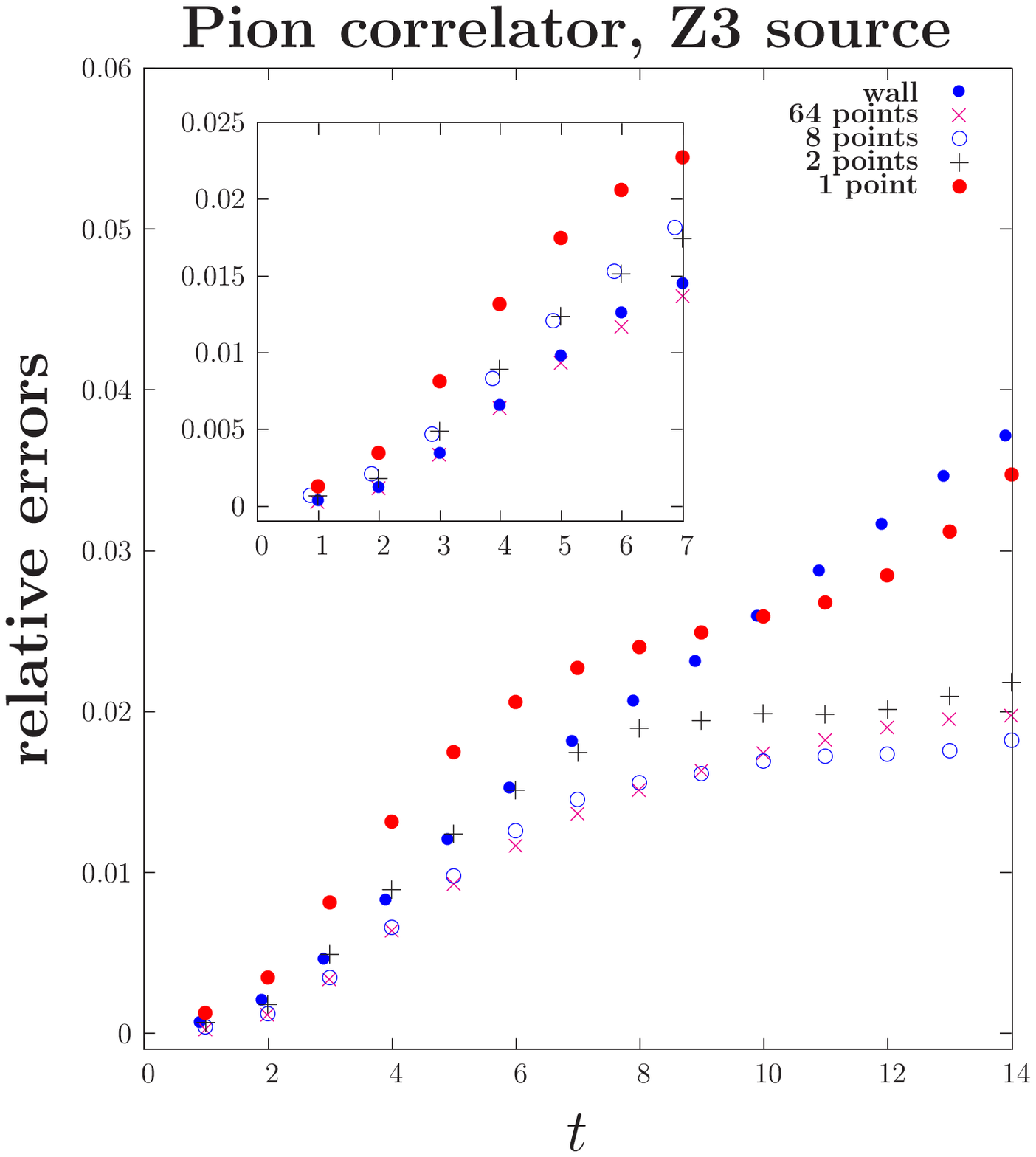}}

	\leavevmode
	\subfigure [ Rho ]
	{ \label{rho_relerr}
	  \includegraphics[width=5cm,height=6cm]{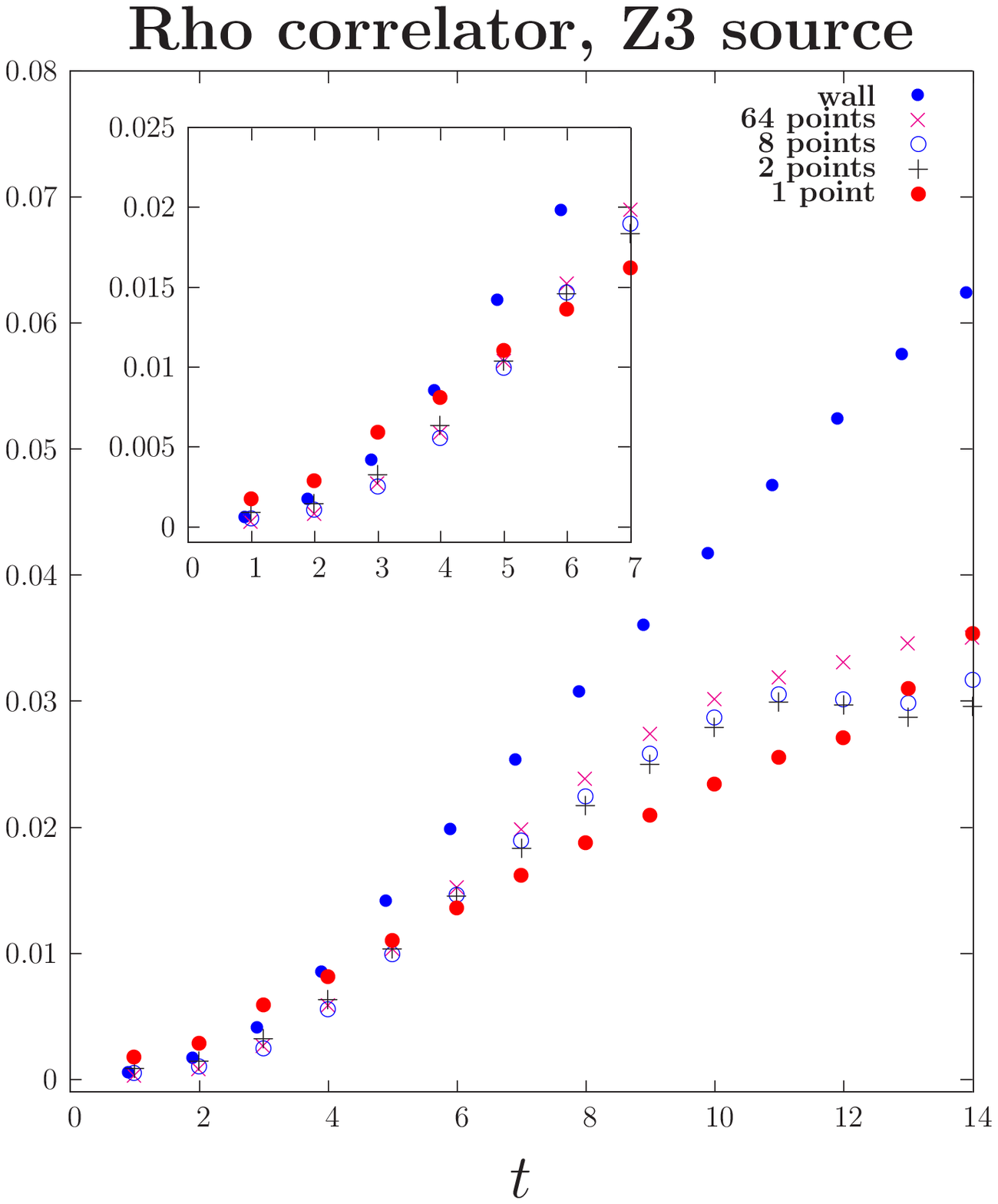}}

	\leavevmode
	\subfigure [ Nucleon ]
	{ \label{nu_relerr}
	  \includegraphics[width=5cm,height=6cm]{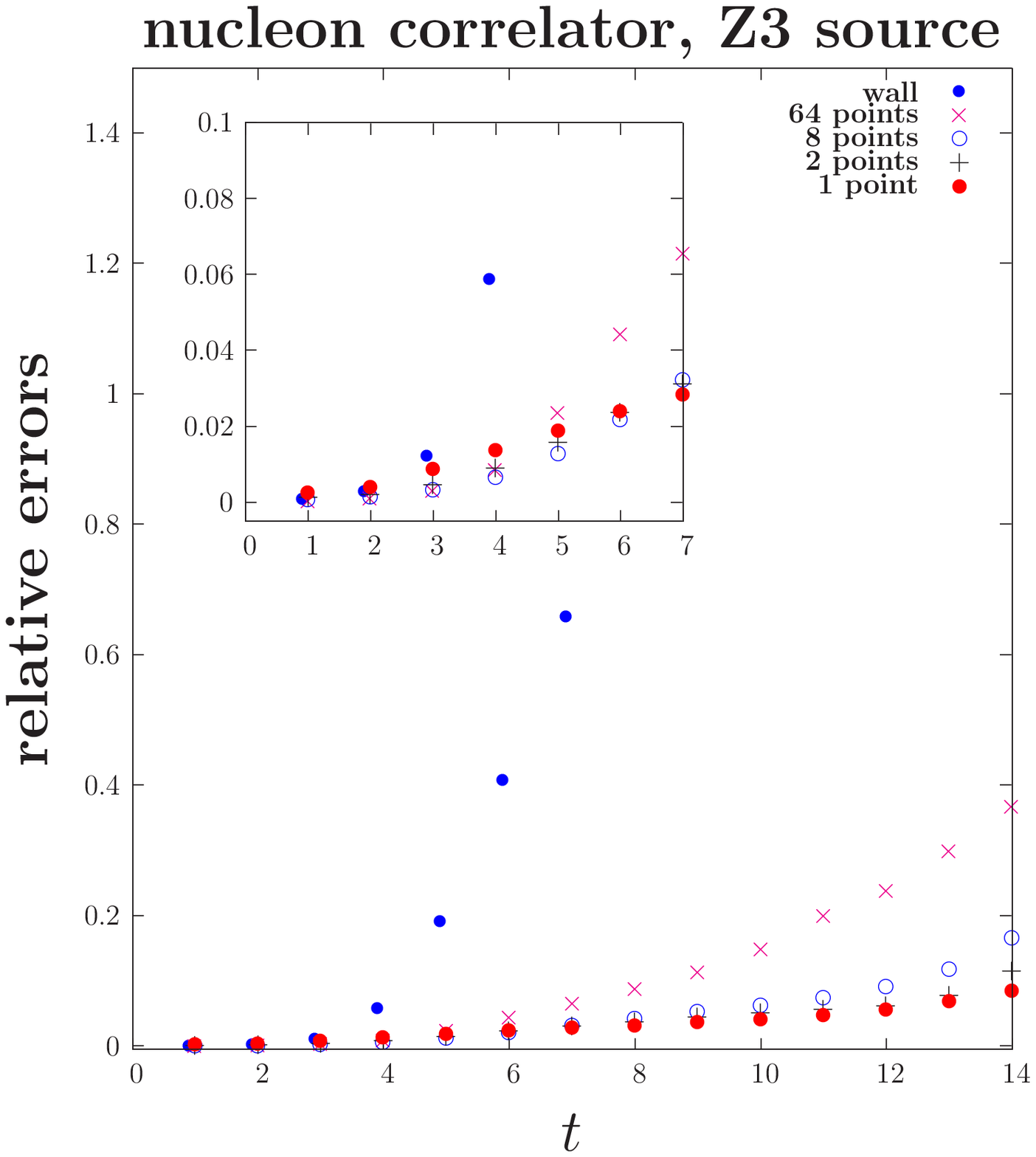}}
}
\caption{Relative errors for (a) pion, (b) rho, and (c) nucleon correlators for different $Z_3$ sources. These are
calculated with Wilson fermion with $\kappa = 0.154$ and $\beta = 6$.}
\label{relerr_Wilson}
\end{center}
\end{figure*}

    To illustrate the above analysis numerically, we show the relative errors of the
pseudoscalar, vector, and nucleon correlators from the Wilson fermion ($\kappa = 0.154$ which corresponds 
to the strange quark mass) on 100 quenched Wilson gauge configurations ($\beta = 6.0$) with one $Z_3$ wall source. 
As shown in Fig.~\ref{relerr_Wilson}, 
the relative errors for the pion (Fig.~\ref{pi_relerr})
is about the same as that for a point source which tends to level off in the range $ t = [7,11]$
as expected. For the $\rho$ (Fig.~\ref{rho_relerr}) and nucleon (Fig.~\ref{nu_relerr}), we observe
that, in addition to the expected rise of noise-to-signal ratio at large $t$, the $Z_3$ wall source
result is worse for the $\rho$ and much worse for the nucleon than those of the point source
({\it{N.B.}} The scales for the ordinates are different in the three sub-graphs.) This
is consistent with the extra $1/\sqrt{V_3}$ factor in Eq.~(\ref{s2n_N}) for the nucleon.

    The undesirable large variance of the noise estimate is rooted in the noise contamination from the
neighboring sites which goes down slowly with $N_n N_g$ in Eq.~(\ref{standard_error}) because $\sigma_n$ is
more than an order of magnitude larger than $\sigma_g$~\cite{dsd09}. To alleviate this difficulty, unbiased
subtraction of contamination from the neighboring sites has been employed to reduce the variance in the
calculation of quark loops~\cite{tdl98,dll95,dll96,dlw98,ddd09}. In the case of connected insertions, it is
found that dilution in time slices~\cite{fjo05,kfh07} is effective in reducing contamination from the nearby
time slices. To carry the suggestion further, dilution of space points within a time slice should
further reduce the variance~\cite{edwards08}. To check this idea, we use a $Z_3$ grid source with support on
certain spatial grid points in a time slice to calculate the quark propagator. 
This we refer to as the $Z_3$ grid source. The results are also included
in Fig.~\ref{relerr_Wilson}. The 64/8 points refers to the points of the grid which are separated by 4/8 lattice
spacings in each spatial direction on a time slice of the $16^3 \times 24$ lattice. We first observe that
they are all better than the $Z_3$ wall source. In the case of pion, they are even better than the point source at
large $t$. For $\rho$ and nucleon, it is interesting to note from the inserts in Fig.~\ref{rho_relerr} and Fig.~\ref{nu_relerr}
that the relative errors of $Z_3$ grids are smaller than those of the point source roughly in the time ranges which are smaller
than the corresponding spatial separations of the grid points. We will return to this point later when we discuss heavy quarks.

\subsubsection{Low-mode substitution (LMS)}

    From the above study, we conclude that the noise grid source itself does not improve the statistics of the hadron
correlators over the point source except for the pseudoscalar meson. Next, we shall consider low-mode substitution. Since the meson
correlators at large time separation are dominated by the low-energy modes, substituting the low-frequency
part of the noise estimated correlator $C_{LL}$  with the exact one $\tilde{C}_{LL}$ from the eigenmodes, as is
outlined in Eq.~(\ref{C_LL}), has been shown to reduce the variance~\cite{gs04,fjo05,kfh07}. However, the
contributions from hadrons on different sites of the source time slice are correlated, particularly among the
nearby neighbors. To see how correlated they are, we plot the relative errors of $\tilde{C}_{LL}$ at large time
separation for the wall source as well as the grid sources with 1, 8, 64, 128, 256, and 512 grid points.

\begin{figure*}[hbtp]
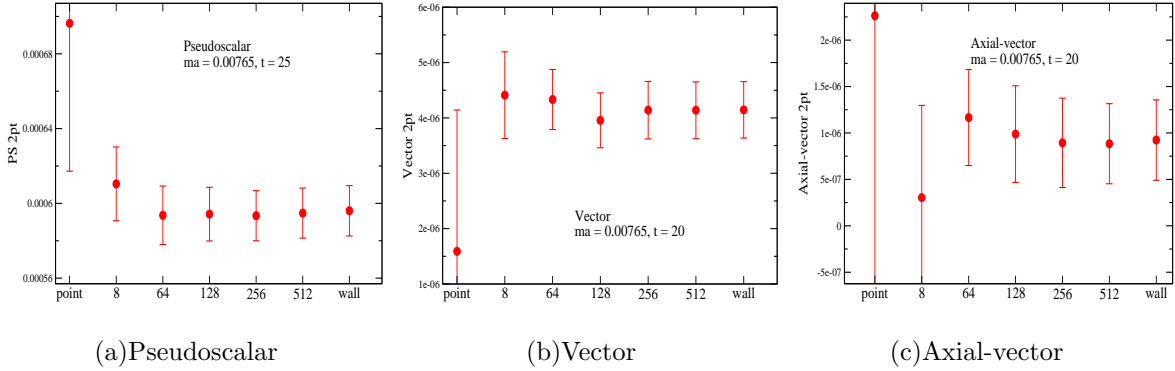

\begin{center}
\mbox{
	\leavevmode
	\subfigure [ Pseudoscalar ]
	{ \label{pi_low_rel}
	  \includegraphics[width=5cm,height=4cm]{figures/ps_m8_t25.eps}}

	\leavevmode
	\subfigure [ Vector ]
	{ \label{rho_low_rel}
	  \includegraphics[width=5cm,height=4cm]{figures/vector_m8_t20.eps}}

	\leavevmode
	\subfigure [ Axial-vector ]
	{ \label{nu_low_rel}
	  \includegraphics[width=5cm,height=4cm]{figures/axial_m8_t20.eps}}
}
\caption{Relative errors from the low-frequency modes at large $t$ separation ($t= 25$ for the
pseudoscalar meson and $t=20$ for the vector and axial-vector mesons) with different number of grid points.}
\label{CLL_errors}
\end{center}
\end{figure*}

These grid points are spaced uniformly in each spatial direction on the source time slice. Plotted in Fig.~\ref{CLL_errors}
are relative errors for the pseudoscalar, vector, and axial-vector mesons at a quark mass which corresponds to a pion mass 
$\sim 200$ MeV.
These are calculated from the 400 pairs of eigenmodes from 50 $32^3 \times 64$ DWF lattice configurations with $m_l = 0.004$.
We see that the relative errors are practically the same from the whole wall down to 64 grid points. This shows that
there is no practical advantage to use the noise wall source, since the low-mode contributions from mesons emerging
from different sites are highly correlated. $\tilde{C}_{LL}$ would be the same with as little as 64 grid points.

\vspace*{1cm}

    We have learned that if the grid points are too dense, such as close to that of the wall, there is large
noise contamination
from the neighboring sites. On the other hand, the low-mode substituted part of the correlator $\tilde{C}_{LL}$ is highly
correlated among neighboring points. Thus, the clear choice is to reduce the wall source to a grid source with an optimal
separation between the grid points to reduce noise contamination and, at the same time, not to sacrifice the variance
reduction from
low-mode substitution and the gain in statistics with multiple grid points. This would be a many-to-all approach as opposed
to the all-to-all approach. It may not make much of difference for the pion, but is expected to work better for
other mesons and the baryons. This optimal choice of grid points on a lattice could depend on the number of eigenmodes and
perhaps the hadrons, such as mesons vs baryons, in addition to the balance between high- and low-modes. We have not done
a detailed analysis in this regard. We shall, nevertheless, present results based on 50 DWF configurations on the $32^3 \times 64$
lattice ($m_l = 0.004$) with a $Z_3$ grid source, which has support on 64 points (4 points in each spatial direction with
8 lattice spacings apart), and low-mode substitution with 400 pairs of low-frequency modes plus the zero modes.

\begin{figure}[htbp]
  \centering
  \subfigure[] 
     {\label{PS_4a}
     {\includegraphics[width=7cm,height=6cm]{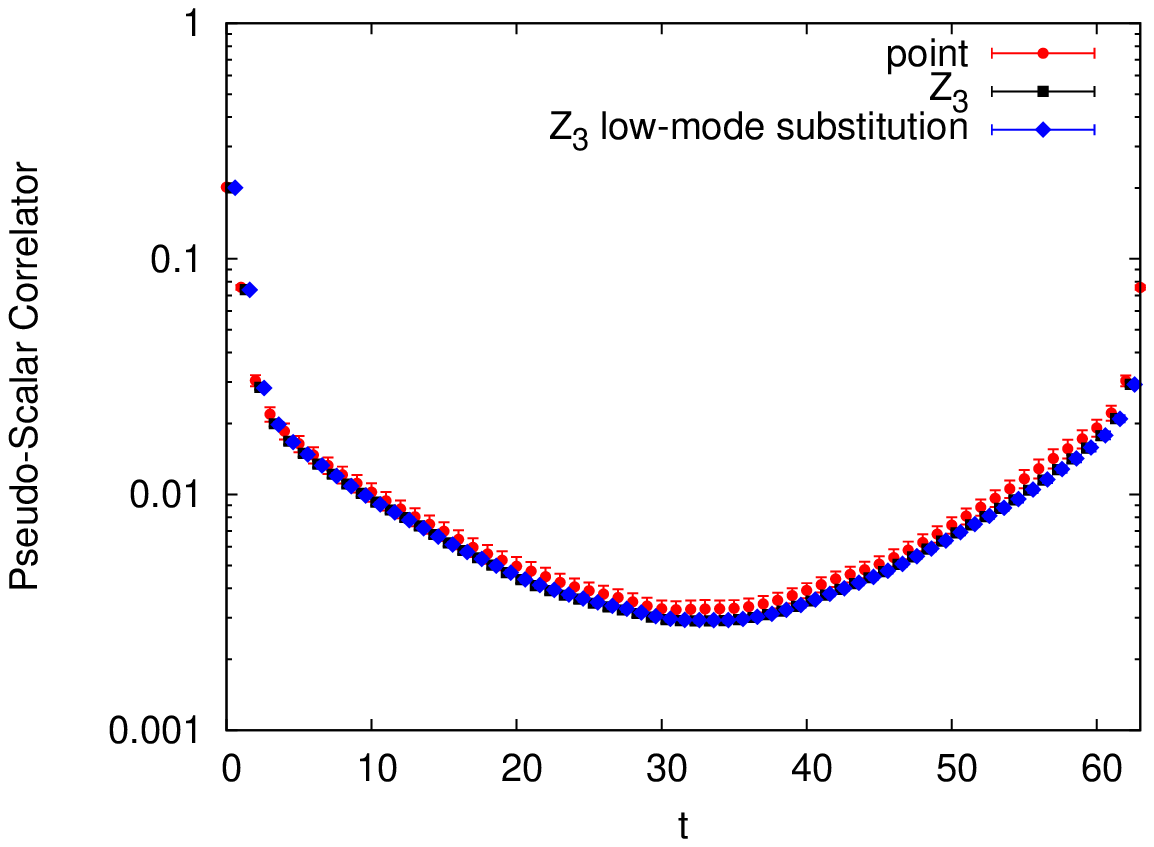}\ \ \ \ }}
  \hspace{0.6cm}
  \subfigure []
     {\label{PS_4b}
     {\includegraphics[width=7cm,height=6cm]{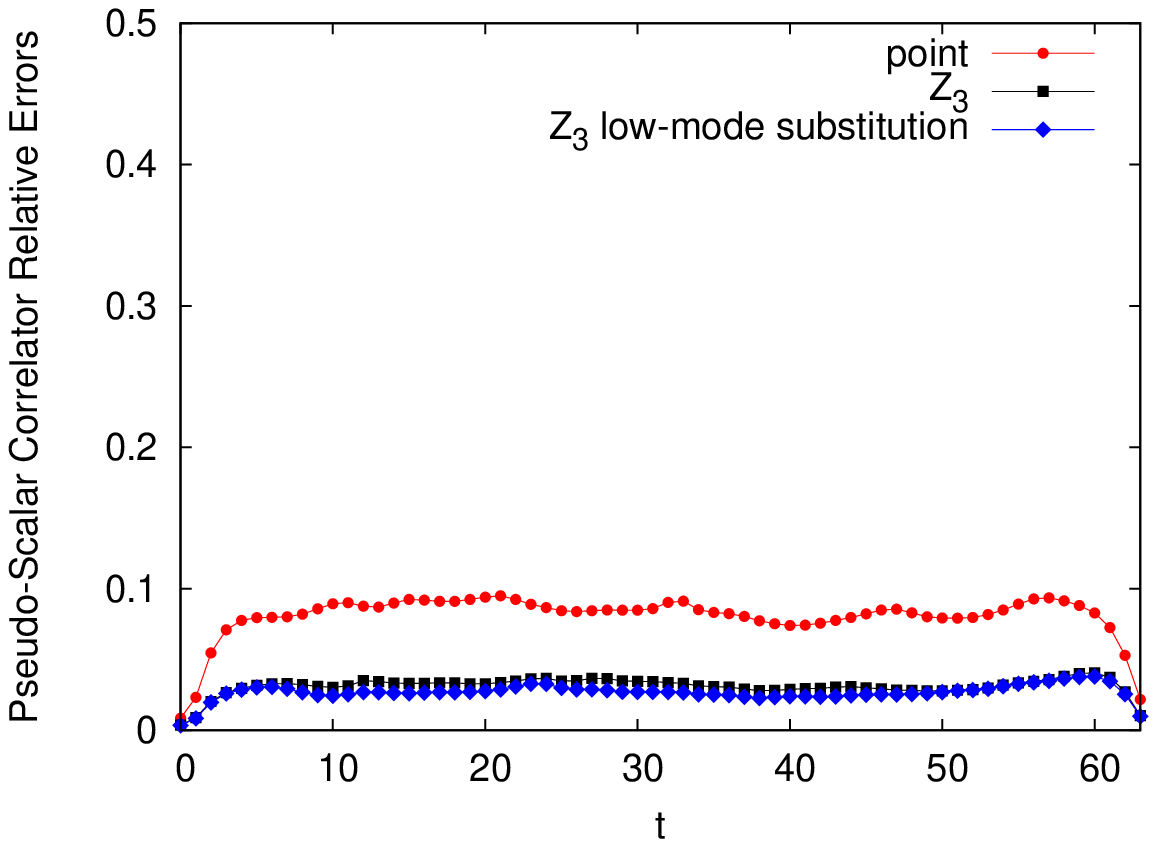}\ \ \ \ }}
  \caption{(color online) (a) The pseudoscalar meson correlators from the point (circle), the $Z_3$ grid source with
64 grid points (square) and the $Z_3$ grid source with low-mode substitution (diamond) are plotted as a function of $t$.
(b) The respective relative errors are plotted as a function of $t$.}
  \label{pion_comp_4}
\end{figure}

    We first plot in Fig.~\ref{PS_4a} the pseudoscalar correlators from the point source, the $Z_3$ grid source
with 64 grid points, and the $Z_3$ grid source with low-mode substitution for the case with pion mass at $\sim 200$ MeV.
Also plotted in Fig.~\ref{PS_4b} are their respective relative errors as a function of $t$. We see that the relative 
errors of the $Z_3$ grid source with or without low-mode substitution
is about a factor of 3 smaller than that of the point source in practically all the time range.

\begin{figure}[htbp]
  \centering
  \subfigure[] 
     {\label{PS_18a}
     {\includegraphics[width=7cm,height=6cm]{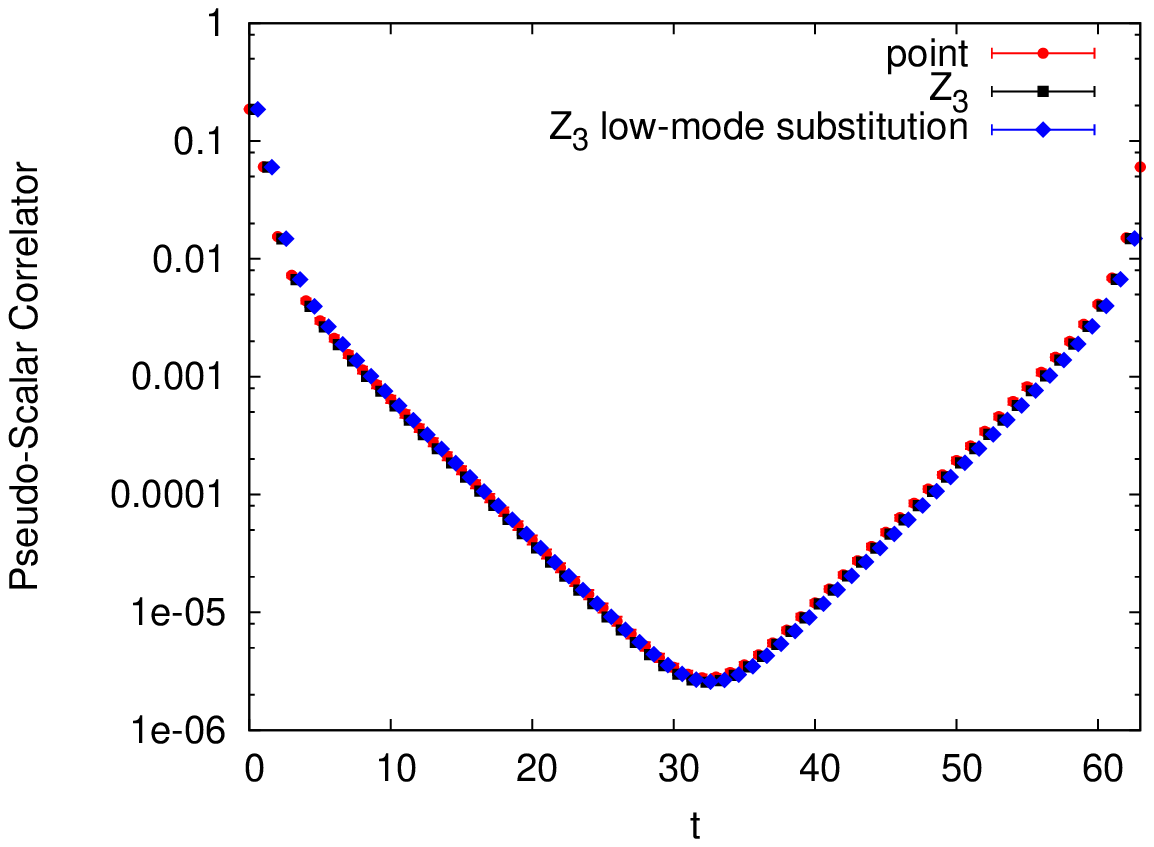}\ \ \ \ }}
  \hspace{0.6cm}
  \subfigure []
     {\label{PS_18b}
     {\includegraphics[width=7cm,height=6cm]{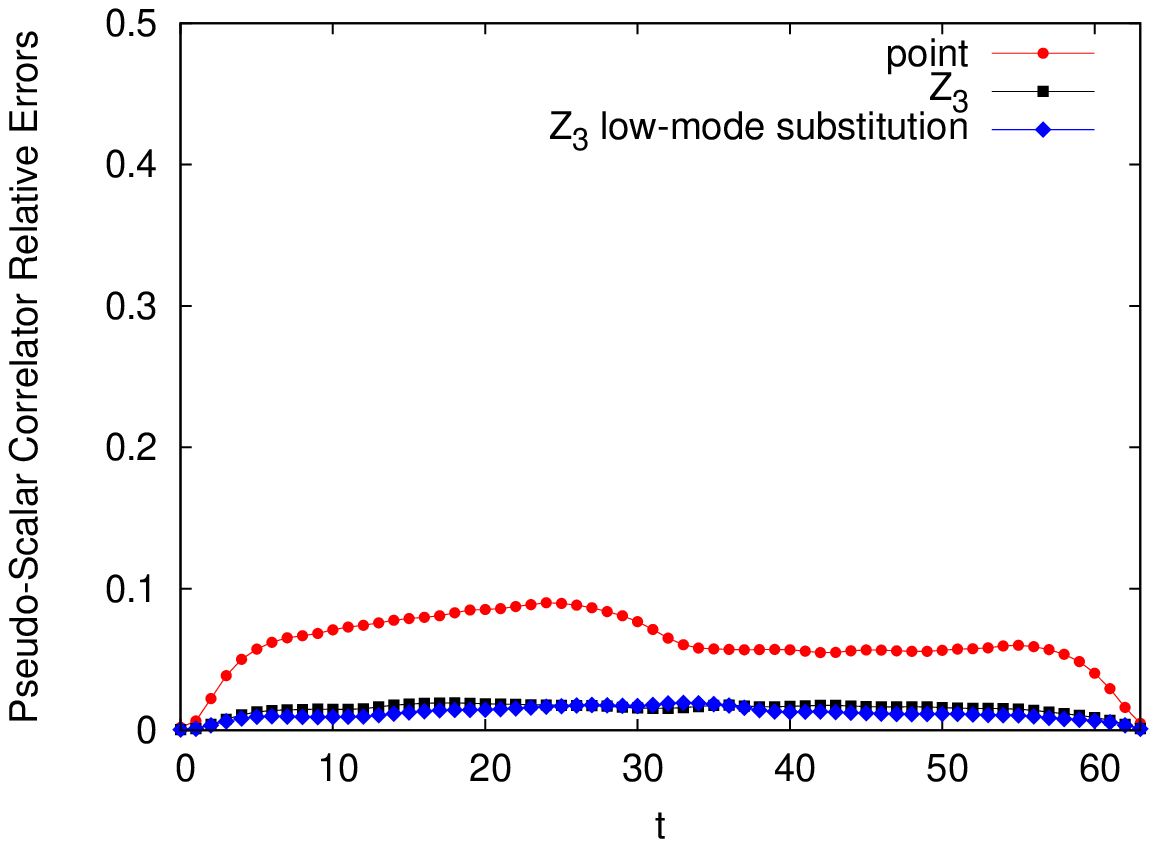}\ \ \ \ }}
  \caption{(color online) The same as Fig.~\ref{pion_comp_4} for the strange quark mass corresponding to pseudoscalar mass at $\sim 670$ MeV.}
  \label{pion_comp_18}
\end{figure}

    A similar situation exists for the strange quark. The results with pseudoscalar mass at $\sim 670$ MeV are
 plotted in Fig.~\ref{pion_comp_18}.

\begin{figure}[htbp]
  \centering
  \subfigure[] 
     {\label{PS_27a}
     {\includegraphics[width=7cm,height=6cm]{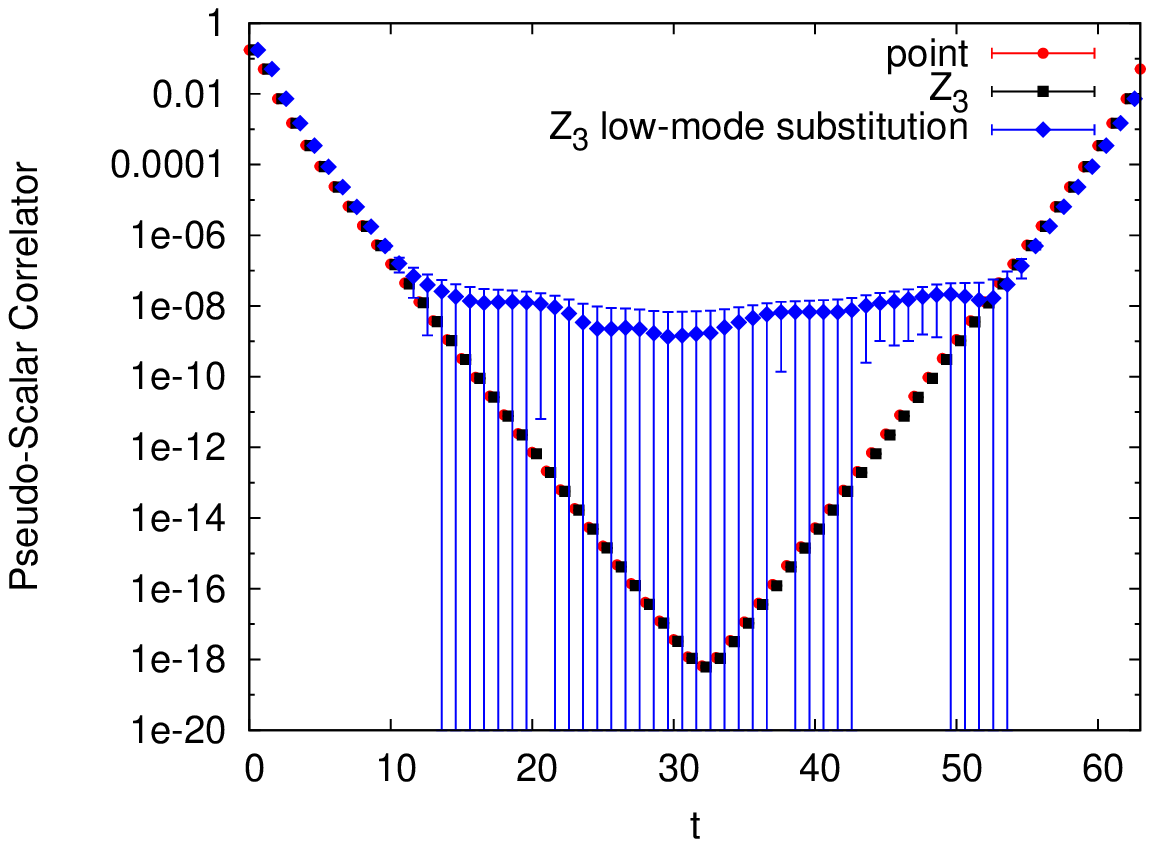}\ \ \ \ }}
  \hspace{0.6cm}
  \subfigure []
     {\label{PS_27b}
     {\includegraphics[width=7cm,height=6cm]{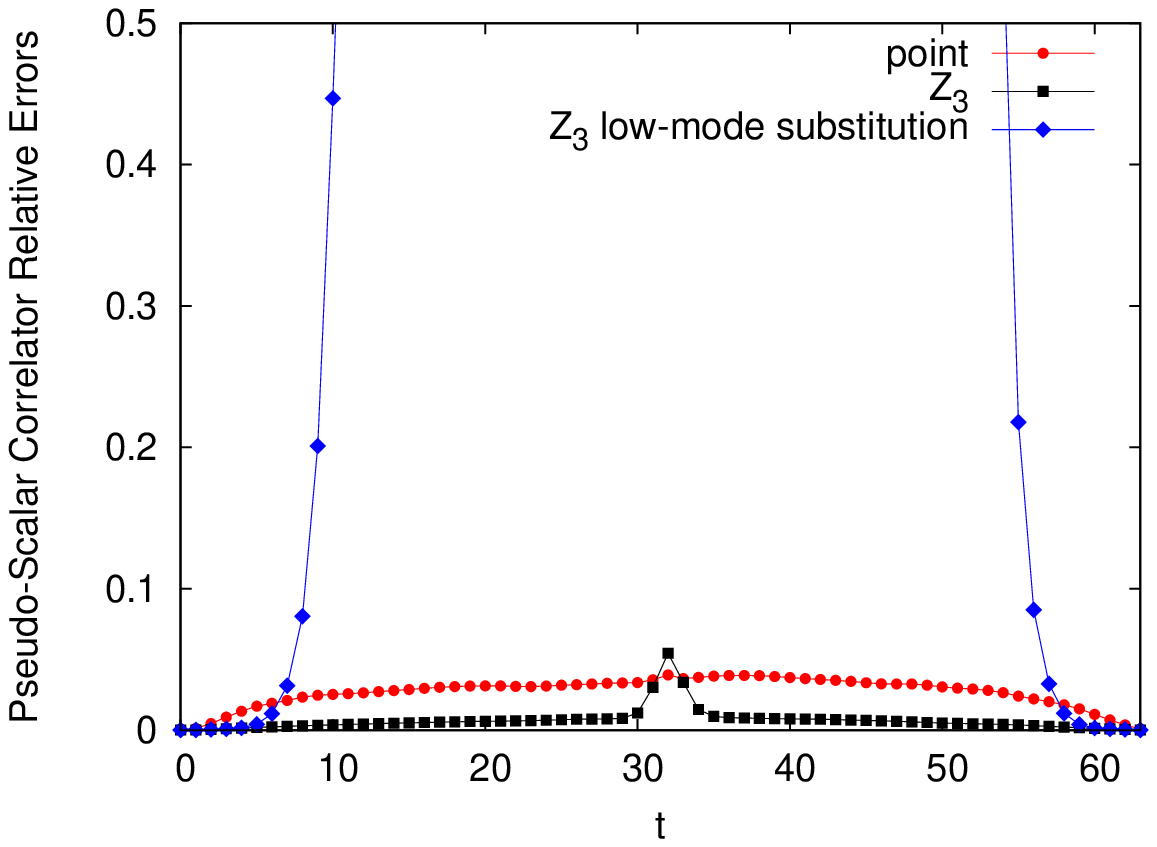}\ \ \ \ }}
  \caption{(color online) The same as Fig.~\ref{pion_comp_4} for the charm quark mass corresponding to pseudoscalar mass at $\sim 2979$ MeV.}
  \label{pion_comp_27}
\end{figure}

    The case for the charm quark is different. We see in Fig.~\ref{PS_27a} for the quark mass around the charm,
that the correlator from the $Z_3$ grid source with low-mode substitution levels off for $t \ge 12$ and its relative
error becomes larger those that of the point and the $Z_3$ grid sources for $t \ge 7$. This is the classic example
where the signal falls off exponentially and the noise estimate levels off at some stage due to a constant variance.
In this case, the results from the low-mode substitution will not be useful. On the other hand, we notice that the
error from the $Z_3$ grid is about a factor of 3 smaller than that of the point source at large $t$ which resembles the
situation with the low-mode substitution for the light quarks when the time separation is less than that of the spatial
separation of the grid points. It is interesting to ponder why this is so. Although
without a proof, we venture to speculate that since the quarks are confined
and the charm quark velocity $v$ is about 30\% of the speed of light, spatial separation
$\Delta x$ of two grid points greater than $v$ times the Euclidean time separation
$\Delta t$, i.e. $\Delta x > v \Delta t$, is `space like' in the Minkowski space sense and that would limit the interference
of the two sources. For the minimal spatial separation of
8 lattice spacings, this limiting $\Delta t$ is 27 which is close to the mid point of the
time extent of 64. This is consistent with what we observed earlier for the light quarks that the
relative errors of $Z_3$ grids are smaller than those of the point source roughly in the time range shorter than the
spatial minimal separation of the grid points. In that case, the
light quarks are expected to propagate close to the speed of light.

\begin{figure}[htbp]
  \centering
  \subfigure[] 
     {\label{V_4a}
     {\includegraphics[width=7cm,height=6cm]{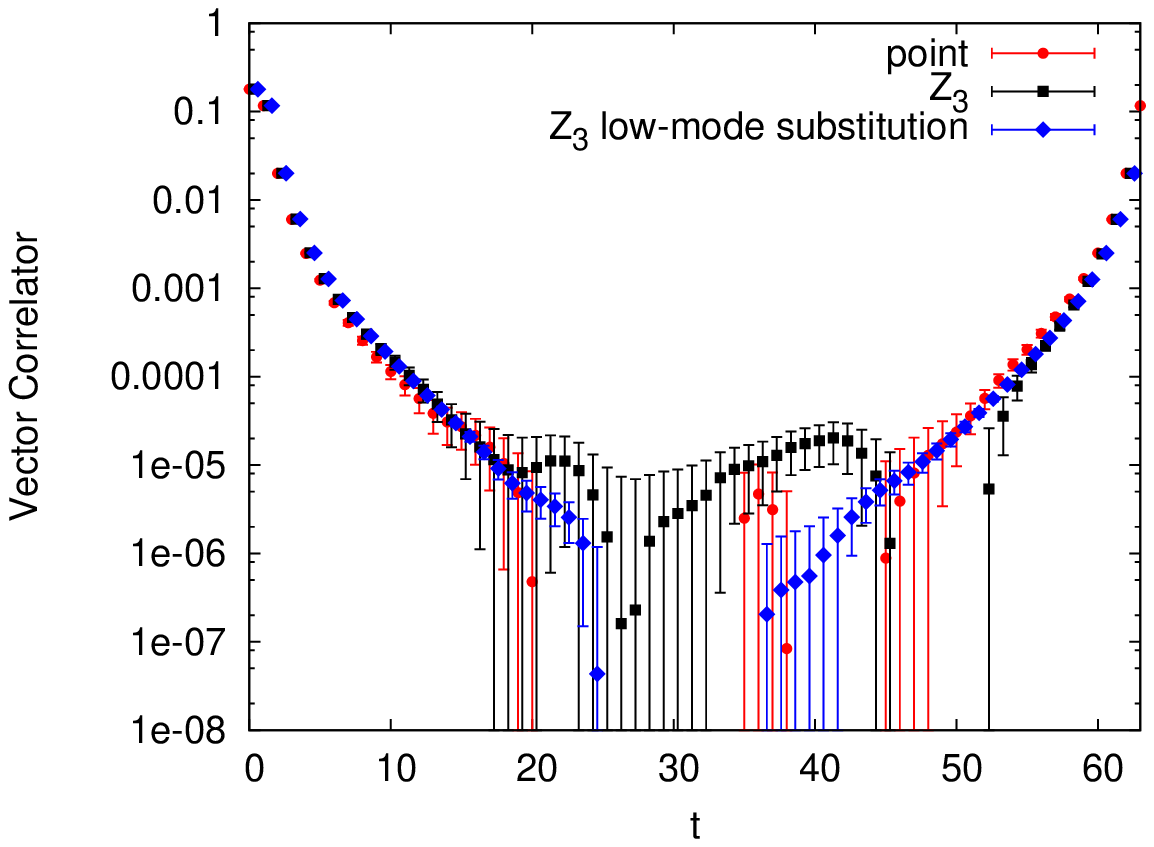}\ \ \ \ }}
  \hspace{0.6cm}
  \subfigure []
     {\label{V_4b}
     {\includegraphics[width=7cm,height=6cm]{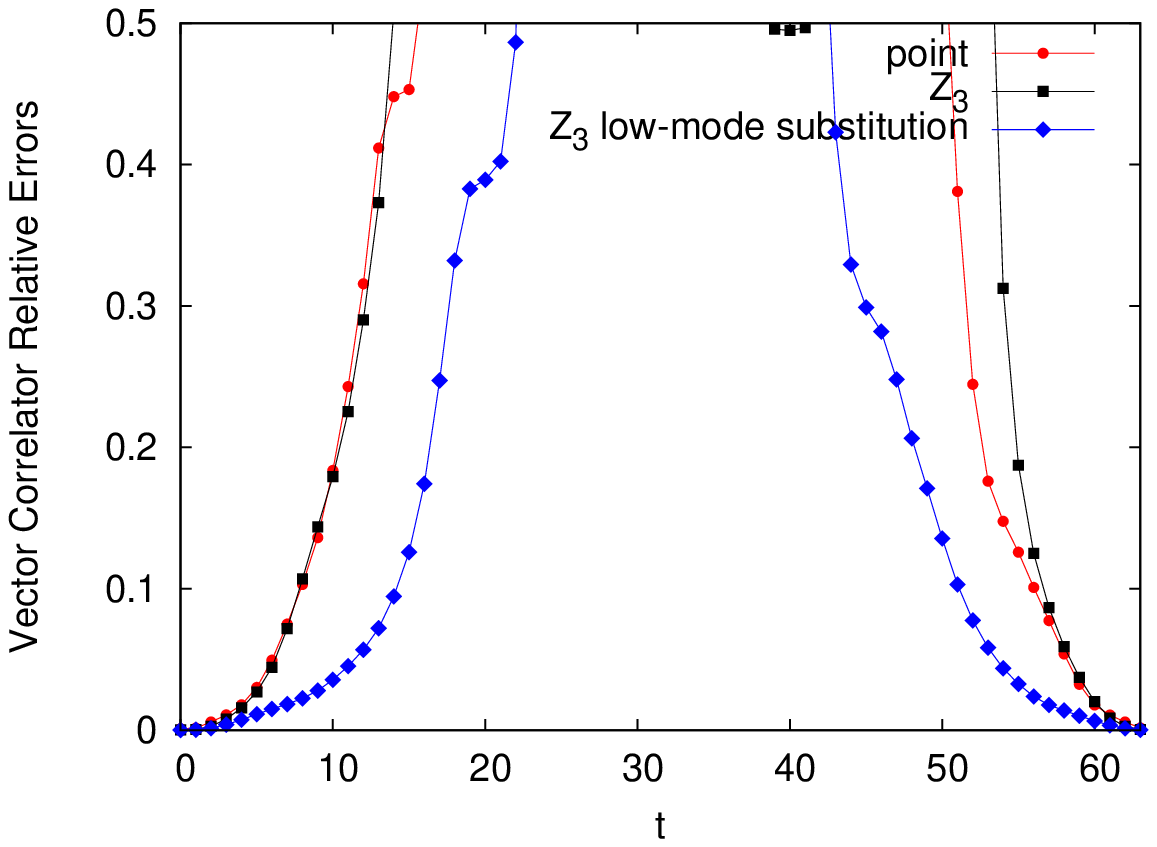}\ \ \ }}
  \caption{(color online) The same as Fig.~\ref{pion_comp_4} for the vector meson correlator with
  light quark ($m_{\pi} \sim 200$ MeV.)}
  \label{vec_comp_4}
\end{figure}

\begin{figure}[htbp]
  \centering
  \subfigure[] 
     {\label{V_18a}
     {\includegraphics[width=7cm,height=6cm]{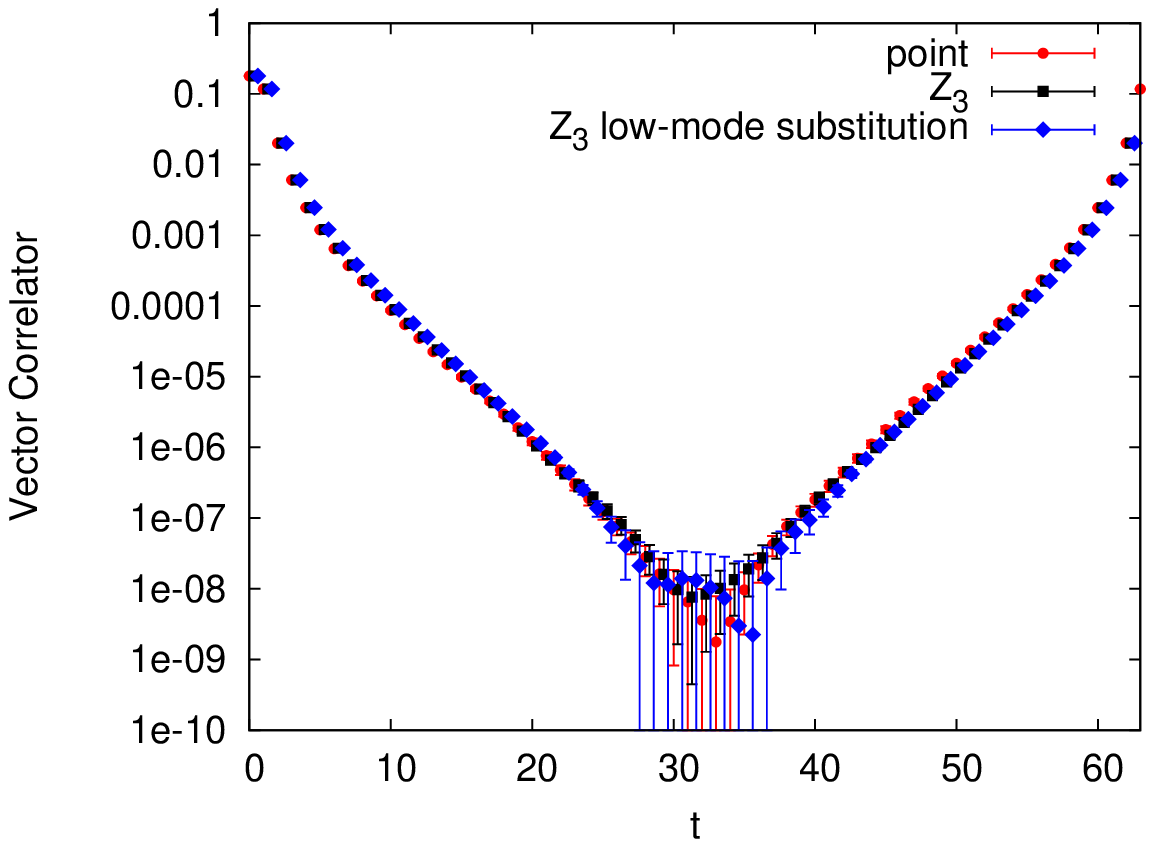}\ \ \ \ }}
  \hspace{0.6cm}
  \subfigure []
     {\label{V_18b}
     {\includegraphics[width=7cm,height=6cm]{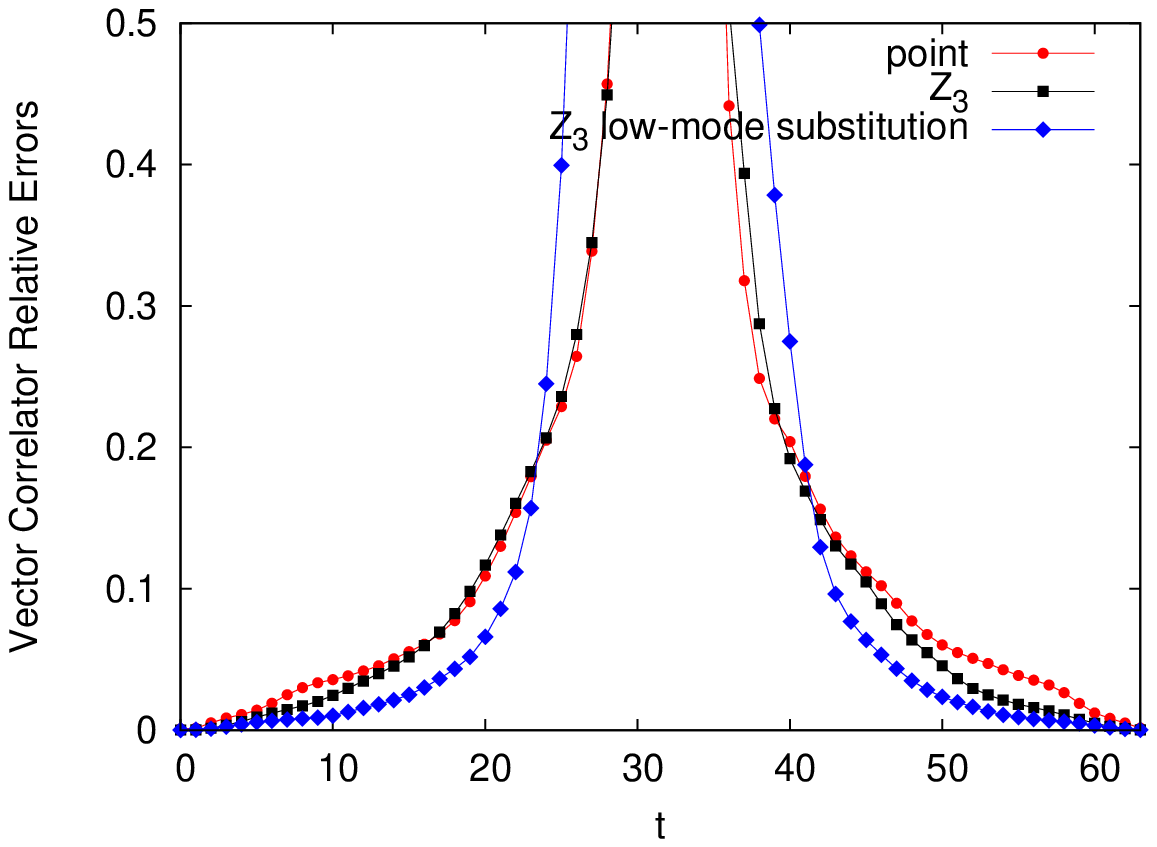}\ \ \ \ }}
  \caption{(color online) The same as Fig.~\ref{pion_comp_18} for the vector meson correlator with strange quark 
mass corresponding to pseudoscalar mass at $\sim 670$ MeV.}
  \label{vec_comp_18}
\end{figure}

\begin{figure}[htbp]
  \centering
  \subfigure[] 
     {\label{V_27a}
     {\includegraphics[width=7cm,height=6cm]{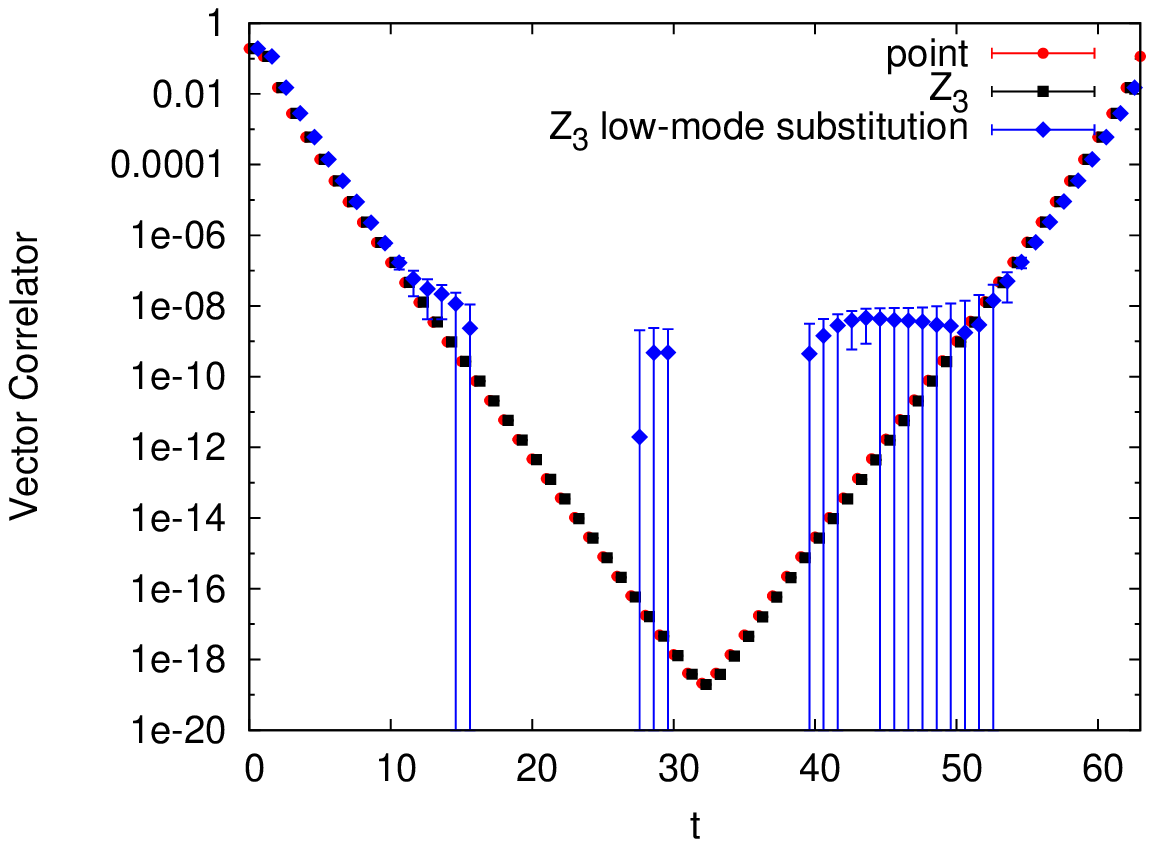}\ \ \ \ }}
  \hspace{0.6cm}
  \subfigure []
     {\label{V_27b}
     {\includegraphics[width=7cm,height=6cm]{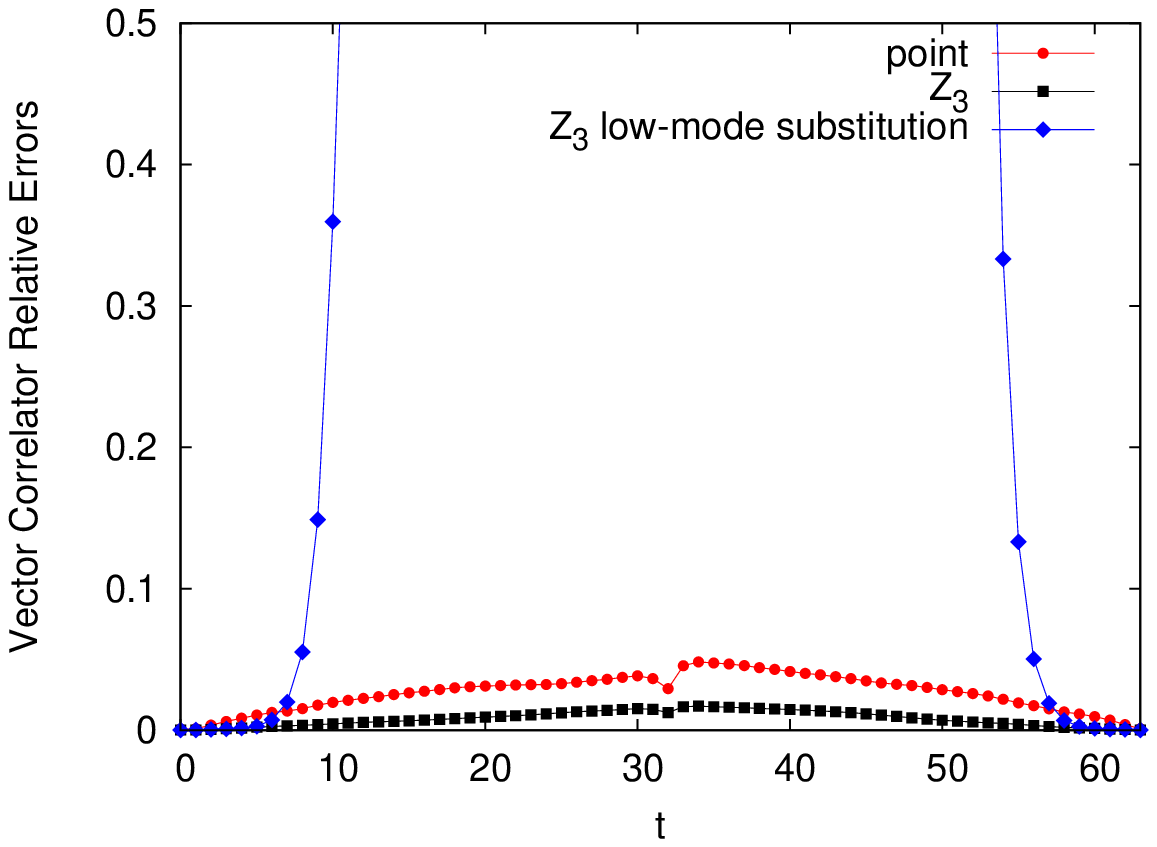}\ \ \ \ }}
  \caption{(color online) The same as Fig.~\ref{pion_comp_27} for the  vector meson correlator with the charm quark 
mass corresponding to pseudoscalar mass at $\sim 2979$ MeV.}
  \label{vec_comp_27}
\end{figure}

   In the case of the vector meson, we see in Fig.~\ref{vec_comp_4} that for light quark mass ($m_{\pi} \sim 200$
MeV), the relative error due to the $Z_3$ grid source with LMS is a factor of 4 to 5 times smaller than those of the
point and the grid sources. For the strange quark mass region (Fig.~\ref{vec_comp_18}), the LMS has smaller
relative error in the range $t < 22$ and then the error becomes larger than those of the point
and the grid sources beyond this range. The charm quark case in Fig.~\ref{vec_comp_27} is similar to that of
the pseudoscalar meson in Fig.~\ref{pion_comp_27}. Although we do not show them here, the axial and scalar
meson correlators are similar to the vector meson case.

\begin{figure}[htbp]
  \centering
  \subfigure[] 
     {\label{N_8a}
     {\includegraphics[width=7cm,height=6cm]{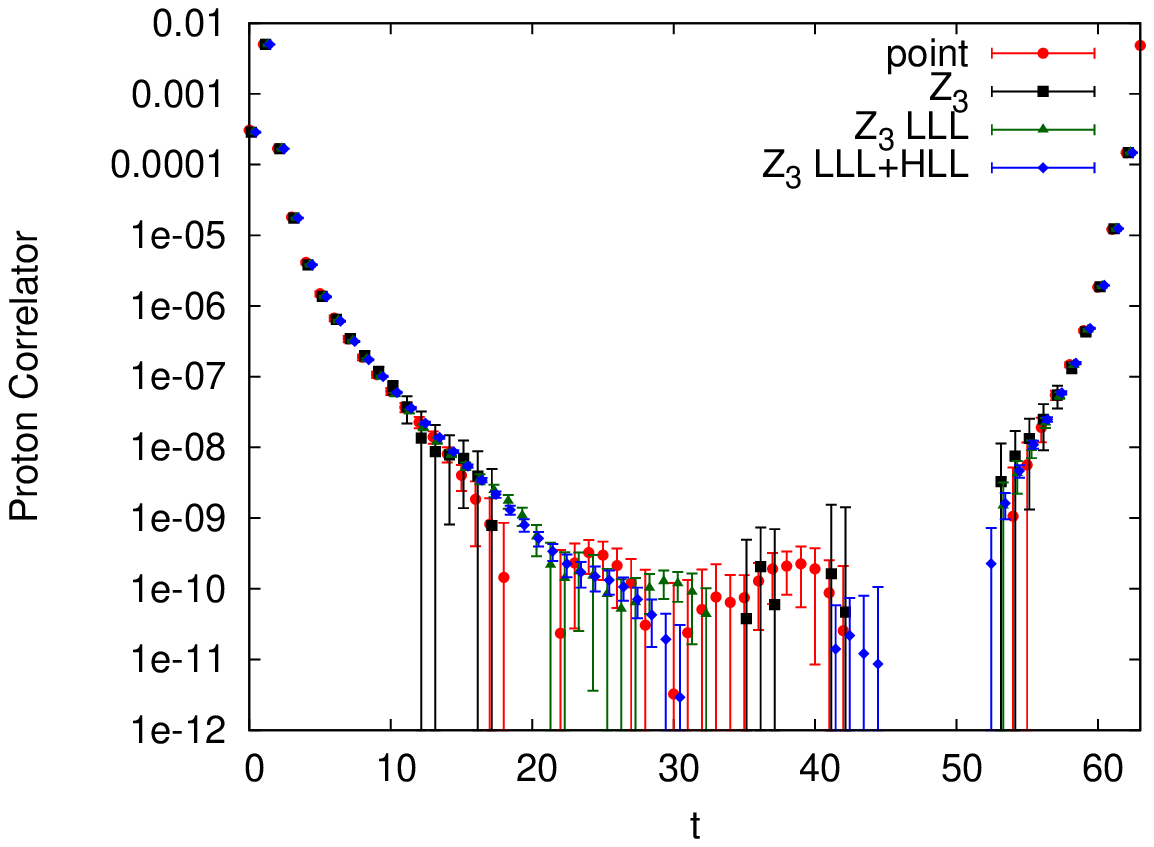}\ \ \ \ }}
  \hspace{0.6cm}
  \subfigure []
     {\label{N_8b}
     {\includegraphics[width=7cm,height=6cm]{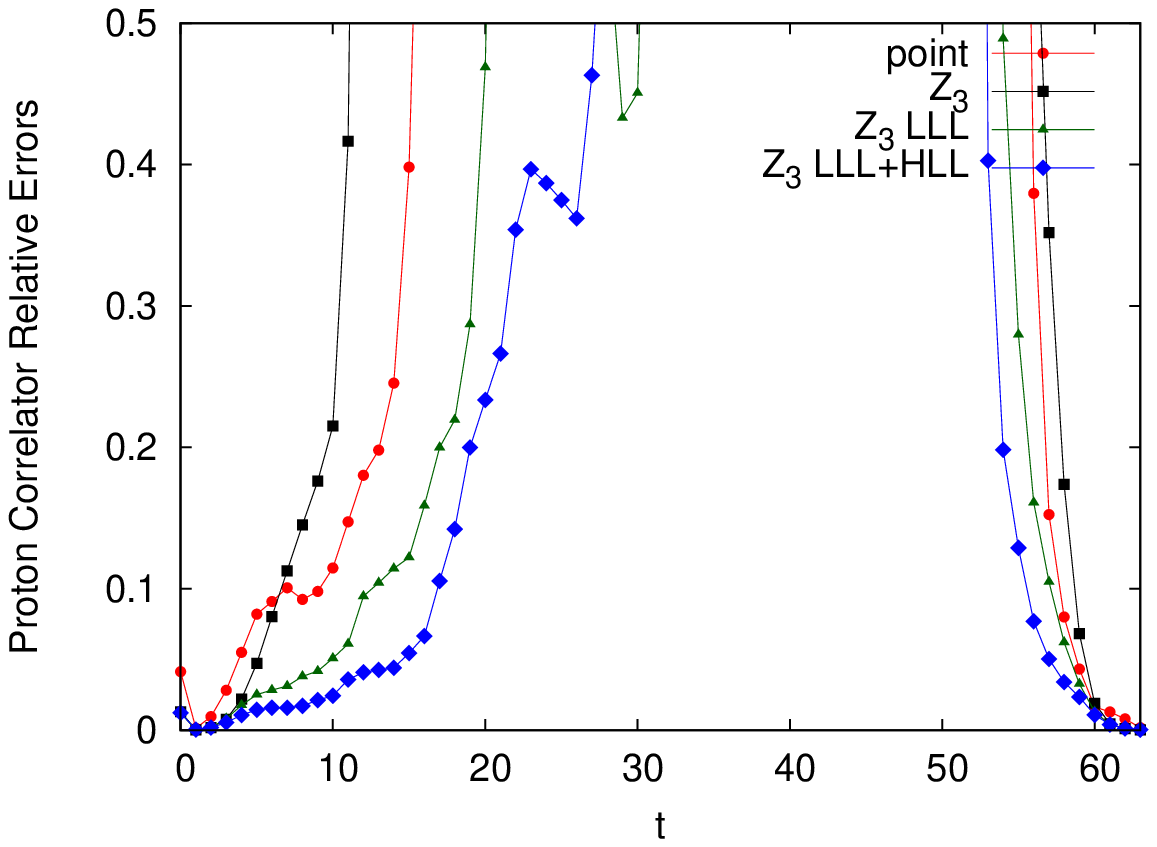}\ \ \ }}
  \caption{(color online) The same as Fig.~\ref{pion_comp_4} for the nucleon correlator with quark
  mass corresponding to pseudoscalar mass at $\sim 300$ MeV.}
  \label{N_comp_8}
\end{figure}

     We plot the results of the nucleon in Fig.~\ref{N_comp_8} for the quark mass which
has a pion mass at $\sim 300$ MeV. The left half of the time range is the nucleon channel and the right half is
the $S_{11}$ channel. For the nucleon, we see that the relative error
of the grid source becomes larger than that of the point source at $t \sim 7$ which is
again close to the 8 lattice spacing separation of the grids. The points labeled by
$Z_3$ LLL are those with LMS for all three quarks. $Z_3$ LLL+HLL represents
those with LMS for two quarks in addition to $Z_3$ LLL. They are defined in Eq.~(\ref{baryon_sub}).
We observe that the error from $Z_3$ LLL is smaller than those of the point and grid sources. This reverses the
situation where the $Z_3$ grid source itself is worse than the point source as we remarked before. With more
LMS from $Z_3$ LLL+HLL, the relative error is further reduced and is brought down below that of the point source 
by more than a factor of 4 at large time separation.

\section{Summary}

   To summarize, we have carried out a study of calculating overlap fermion propagators and
hadron correlators on the $2+1$ flavor domain wall fermion configurations on $16^3 \times 32, 24^3 \times 64$, and
$32^3 \times 64$ lattices with both deflation in the inversion and
low-mode substitution in constructing the correlators.

   With HYP smearing and low-mode deflation, we find a speed up from
$\sim$ 23 for the $16^3 \times 32$ lattice with 200 pairs of eigenmodes and $sim$ 51 for the $24^3 \times 64$ lattice with
200 pairs of eigenmodes to $\sim 79$ for the larger $32^3 \times 64$ lattice with 400 pairs of eigenmodes. 
The cost of the overhead for calculating eigenmodes is 4.5, 4.9 and 7.9 propagators for the above lattices, respectively, which
will be amortized with calculation of propagators for more sources. We have calculated the
quark mass dependence of the hyperfine splitting and find that one can accommodate charm
quarks with small $O(m^2a^2)$ error. Since this is a mixed action approach with overlap on DWF sea, we use the finite
volume boundary condition property of the scalar correlator to estimate
the low-energy constant $\Delta_{mix}$ for finite lattice spacing which is needed for
the mixed action partially quenched chiral perturbation theory extrapolation to the physical
point and the continuum limit. The preliminary result of $\Delta_{mix} \sim (427 {\rm MeV})^4$ turns out to be small.
It only shifts the 300 MeV mixed valence-sea pion mass by $\sim 10$ MeV at $a^{-1} = 2.32$ GeV for the $32^3 \times 64$
lattice and $\sim 19$ MeV at $a^{-1} = 1.73$ GeV for the $24^3 \times 64$ lattice.

    We have examined the signal-to-noise issue for the connected hadron correlators
from the noise source on a time slice and found that the noise wall source is worse than the point source for
all mesons except the pion. It is worse still for the baryon (and multi-quark systems) by a $\sqrt{V_3}$ factor
where $V_3$ is the 3-volume of the time slice. The situation can be ameliorated by reducing the contamination
from neighboring sites with less source points. This introduces the idea of a noise grid source with support
on some uniformly spaced grid points on a time slice. On the other hand, we find that the low-frequency part of
the multiple hadron source with exact eigenmodes is highly correlated so that, beyond 64 grid points on
a time slice of the $32^3 \times 64$ lattice, the relative errors of the meson correlators do not decrease.
These observations led to a suggestion of a new algorithm for the grid noise with low-mode substitution to reduce the 
variance from noise contamination while addressing the low-mode correlation at the same time. 

    We decide to use 64 $Z_3$ grid noise and low-mode substitution with 400 pairs of eigenmodes
on the $32^3 \times 64$ lattice with the light sea mass $m_l = 0.004$ to calculate both the meson and baryon
correlators. We find that for light quarks (pion masses at 200 - 300 MeV), the errors of the mesons and nucleon
masses can be reduced by a factor of $\sim$ 3 to 4 as compared to the point source. In the strange quark region,
the statistical errors of the pion and nucleon masses can be improved by a factor $\sim 3$, but it is not much
improved for the vector meson. We find that the results from low-mode substitution start to degrade beyond
the strange quark region. This is due to the fact that the signal falls off quickly at large $t$ and yet the variance 
of the noise estimation of the high-frequency and the mixed high- and low-frequency parts of the correlator does not fall.
Luckily, the $Z_3$ grid results are still better than the the point source and can reduce the errors of
the charmonium masses by a factor of $\sim 3$. One can use it to address the hadrons involving the charm quark. We should point 
out that the interplay between the noise grid source and low-mode substitution is quite general and is not restricted to a 
particular fermion action.

    So far the study of two-point functions has been favorable. The three-point function with $Z_3$ grid
source and low-mode substitution will undoubtedly pose a different set of challenges.

\section{Acknowledgment}
This work is partially support by U.S. DOE Grants No. DE-FG05-84ER40154, No. DE-FG02-95ER40907 and 
No. DE-FG02-05ER41368. AA is supported by the George Washington University IMPACT initiative. YC is supported by 
Chinese NSFC-Grant No. 10835002. TD is supported in part by Grant-in-Aid 
for JSPS Fellows 21$\cdot$5985. NM is supported in part by DST-SR/S2/RJN-19/2007 fellowhisp. JBZ is supported by Chinese 
NSFC-Grant No. 10675101 and 10835002. The numerical work 
was performed on Franklin at NERSC as well as on Kraken at NICS under the TeraGrid allocation. We thank R. Edwards, B. Joo, 
and T. Kennedy for useful discussion. A. Alexandru, A. Li and K.F. Liu would like to acknowledge the hospitality of 
the Kavli Institute of Theoretical Physics in Beijing where part of this work was carried out during a lattice workshop 
in July 2009.

\end{document}